\documentclass[reprint,aps,prd,superscriptaddress,showpacs,nofootinbib,eqsecnum,amsfonts,amsmath,twocolumn]{revtex4-1}
\usepackage{epsfig}
\usepackage{longtable}

\def\nn{\nonumber}
\def\beq{\begin{equation}}
\def\eeq{\end{equation}}
\def\be{\begin{equation}}
\def\ee{\end{equation}}
\def\bea{\begin{eqnarray}}
\def\eea{\end{eqnarray}}

\def\la{\langle}
\def\ra{\rangle}
\def\ms{\overline{MS}}
\def\ov{\overline}
\def\t{\tilde }
\def\Lam{\Lambda}
\def\eps{\epsilon}
\def\ga{\gamma}
\def\ds{\displaystyle}
\newcommand{\qq}{\langle \bar q q \rangle} 
\newcommand{\lsim}{\raisebox{-0.13cm}{~\shortstack{$<$ \\[-0.07cm] $\sim$}}~}
\newcommand{\gsim}{\raisebox{-0.13cm}{~\shortstack{$>$ \\[-0.07cm] $\sim$}}~}

\begin{document}

\title{$\alpha_S$ from $F_\pi$ and Renormalization Group
Optimized Perturbation}

\author{Jean-Lo\"{\i}c Kneur}
\author{Andr\'e Neveu}
\affiliation{CNRS, Laboratoire Charles Coulomb UMR 5221, F-34095, Montpellier, France}
\affiliation{Universit\'e Montpellier 2, Laboratoire Charles Coulomb UMR 5221, F-34095, Montpellier, France}

\begin{abstract}
A variant of variationally optimized perturbation, incorporating renormalization group properties in a straightforward way,  uniquely fixes 
the variational mass interpolation in terms of the anomalous mass dimension. It is used at three successive orders to calculate the nonperturbative ratio
$F_\pi/\overline \Lambda$ of the pion decay constant and the basic QCD scale in the $\ms$ scheme. 
We demonstrate the good stability and (empirical) convergence properties of this modified
perturbative series for this quantity, and provide simple and generic cures to previous problems of the method, principally the generally
non-unique and non-real optimal solutions beyond lowest order. Using the experimental $F_\pi$ input value we determine 
$\overline\Lam^{n_f=2}\simeq 359^{+38}_{-25} \pm 5$ MeV and $\overline\Lam^{n_f=3}=317^{+14}_{-7} \pm 13$ MeV, where the first quoted errors are 
our estimate of theoretical uncertainties of the method, which we consider conservative.
The second uncertainties come from the present uncertainties in $F_\pi/F$ and $F_\pi/F_0$, where $F$ ($F_0$) is $F_\pi$ in the exact chiral 
$SU(2)$ ($SU(3)$) limits. Combining the $\ov\Lam^{n_f=3}$ results with a standard perturbative evolution  provides a new independent  
determination of the strong coupling constant at various relevant scales, in particular $\overline\alpha_S (m_Z) =0.1174 ^{+.0010}_{-.0005} \pm .001 \pm .0005_{evol}$ 
and $\overline\alpha_S^{n_f=3}(m_\tau)=  0.308 ^{+.007}_{-.004} \pm .007 \pm .002_{evol}$. A less conservative interpretation of our prescriptions favors central values
closer to the upper limits of the first uncertainties.  
The theoretical accuracy is well comparable to the most precise recent {\em single} determinations of $\alpha_S$, 
including some very recent lattice simulation determinations with fully dynamical quarks.
\end{abstract}
\maketitle
\section{Introduction}
In the massless quarks, chiral symmetric, limit of QCD, the strong coupling 
$\alpha_S(\mu)$ at some reference scale $\mu$
is the only parameter. Equivalently the Renormalization-Group (RG) invariant scale
\be
\ov\Lam^{n_f}\equiv \mu e^{-\frac{1}{\beta_0\alpha_S}} (\beta_0\alpha_S)^{-\frac{\beta_1}{2\beta^2_0}}\:(1+\cdots)\;,
\label{Lamdef1}
\ee
in a specified renormalization scheme~\footnote{In (\ref{Lamdef1}) $\beta_0$, $\beta_1$ are
(scheme-independent) one- and two-loop RG beta function coefficients, and ellipsis denote 
higher orders scheme-dependent RG corrections as will be specified below.}, is the fundamental QCD scale. 
As indicated in Eq.~(\ref{Lamdef1}) $\ov\Lambda^{n_f}$ also depends on the number of active quark flavors $n_f$,  
with non-trivial  (perturbative) matching relations at the quark mass thresholds~\cite{PDG,matching4l}.
Values of $\ov\alpha_S(\mu)$ in the $\ms$ scheme have been extracted at various scales from many different 
observables confronted with theoretical predictions, and its present World average is 
impressively accurate~\cite{PDG}: 
\be
\ov\alpha_S(m_Z)= 0.1184\pm 0.0007
\label{asWa}
\ee
However, this averaged value with combined uncertainties is largely
dominated by a combination of a certain class of lattice results~\cite{asHPQCD,asLattpert}: $\ov\alpha_S(m_Z)\simeq .1185\pm .0007$, with central value and uncertainty 
that coincide both with the 2010 and the present World average. So (\ref{asWa}) actually hides much 
larger departures among {\em single} determinations from different methods and observables, most having uncertainties rather in the range $.0016-.004$. 
A statistical combination, however sophisticated, of all those different determinations, remains  difficult. 
Indeed there are long-standing tensions with 
values extracted from deep inelastic scattering~\cite{alphaS_DIS}: $\ov\alpha_S(m_Z)\simeq .114\pm.002$, and even recently~\cite{alphaS_DIS_low}:
$\ov\alpha_S(m_Z)\simeq .1134\pm.0011$. \\
Over the last years, determinations of $\alpha_S$ from various different methods based on lattice simulations  
have been the subject of intense activities.  
One may actually distinguish two rather different classes of lattice determinations of $\alpha_S$: the first very precise one mentioned above~\cite{asHPQCD,asLattpert}, 
is essentially based on calculating {\em short distance} quantities (heavy quark current correlators, Wilson loops, etc) from lattice simulations, 
matched to perturbative evaluations, thus with direct access to $\alpha_S$ in the perturbative regime. 
The second class is rather based on nonperturbative calculations of the basic QCD scale $\ov\Lam$,  
(see e.g. ~\cite{LamlattSchroed,LamlattSchroednew, LamlattWilson,Lamlatttwisted,LamlattVstat,Lam4latt}). For both classes
much progress was made recently with the advent of fully dynamical quark simulations for $n_f\ge 2$ flavors,   
 and the present accuracies achieved by the various lattice determinations are overall impressive~\cite{Lamlattrev}. However, some differences  
and systematic uncertainties remain, due to the needed non-trivial extrapolation to the chiral limit, using input from Chiral Perturbation Theory~\cite{chpt},
and from matching pure lattice results to $\alpha_S(m_Z)$, in the treatment of the truncation of perturbation theory and in     
different assumptions on nonperturbative (power) correction contributions.

It is thus of interest to provide further independent determinations of  $\ov\Lam$ and $\alpha_S$ 
from other observables, and also from other theoretical methods, especially if possible to access
the infrared, nonperturbative QCD regime for $n_f=2$, where a perturbative extrapolation
from $\alpha_S(m_Z)$ is unreliable. In the present paper we pursue our previous exploration~\cite{rgoptqcd1} of a different route to 
estimate such quantities. As will be clear below, 
our approach is more logically to be compared with the above $\ov\Lam$ lattice determinations. Our calculation builds up on a standard perturbative result, but modified in a way able
to grasp the nonperturbative phenomenon of dynamical chiral symmetry breaking (DCSB) due to the light $u$, $d$ (and $s$) quarks. 
The main order parameters of chiral symmetry breaking, namely the  pion decay
constant $F_\pi$ and the chiral quark condensate $\qq$, should be entirely determined by the unique scale 
$\ov\Lam$ in the strict chiral limit, and $F_\pi$ at least is unambiguously and precisely known experimentally. 
However, conventional wisdom usually considers the above intrinsically nonperturbative parameters not calculable in any ways from
perturbative QCD, so that one has to appeal to truly nonperturbative methods like lattice simulations. 
Perhaps the most intuitive reason is the expected nonperturbative regime at the relevant DCSB scale close to $\ov\Lam$,   
implying {\it a priori} large $\alpha_S$ values 
invalidating reliable perturbative expansions. Another often mentioned argument in the literature, typically for the chiral condensate, 
is that the standard QCD perturbative series at arbitrary orders 
are anyhow proportional to the (light) quark masses $m_q$, $q=u,d,s$ (up to powers of $\ln m_q$), 
so trivially vanish in the strict chiral limit $m_q\to 0$. On top of these, other more sophisticated and generally accepted arguments, 
related to the problem of resumming presumed factorially divergent 
perturbative series at large orders\cite{gzj_lopt,renormalons}, 
seem at first sight to further invalidate any perturbative approach to calculate the order parameters. 
We will see that at least the first two arguments above can be circumvented by a peculiar 
modification of the ordinary perturbative expansion in $\alpha_S$, the so-called optimized perturbation (OPT), which may be viewed
in a precise sense as performing a much more efficient use of the purely perturbative information. 
Our recent version of the method essentially supplements the OPT consistently with renormalization group RG properties~\cite{rgopt1,rgoptqcd1} in a
straightforward way, that also appear to give substantial improvements of the convergence properties of the OPT modified series, as is supported by comparison  
with exact results in models simpler than QCD and empirically exhibited in QCD. \\
Our more general goal is thus to improve the RGOPT approach of \cite{rgopt1,rgoptqcd1} to  possibly determine with a realistic accuracy 
some of the chiral symmetry order parameters in QCD or similar quantities in other models. We will propose a simple cure to the principal issues 
encountered previously with non-real optimization solutions, and also clarify  
a number of previously conjectured general features of the RGOPT, which we believe provides now a well-defined and relatively simple prescription to deal with
renormalizable asymptotically free models. \\

We concentrate here on $F_\pi/\ov\Lam$ because it is connected to one of the best known QCD perturbative series at present, up to four-loop order.
This gives us an opportunity to study the eventual stability/convergence properties of the OPT 
at three successive non-trivial orders, in a full QCD context. As a by-product our result provides a new independent determination of 
$\ov\Lam$ and $\alpha_S(\mu)$. \\
Our method has been applied previously in \cite{rgopt1} for
the $D=2$ Gross-Neveu (GN) $O(2N)$ model\cite{GN}, which shares many properties with $D=4$ QCD: it is 
asymptotically free and the (discrete) chiral symmetry is  
dynamically broken with a fermion mass gap. The exact mass gap is known for arbitrary $N$,   
from the Thermodynamic Bethe Ansatz\cite{TBAGN}, allowing accurate tests of the method. Using only the two-loop ordinary
perturbative information, we obtained approximations to the exact mass gap at the percent
or less level~\cite{rgopt1}, for any $N$ values. Moreover, in the simpler large-$N$ limit, the convergence is maximally fast since
the RGOPT gives the exact result already at first order and at all higher orders. These results not only give us some confidence on the reliability 
of the method, but the analogy goes further, as we will explain, since the main properties which can be confronted with exact results in 
the GN model are very similar for the relevant QCD quantities we consider here. \\  

The paper is organized as follows. In Sec. II we review the main features of the OPT and our RGOPT version 
incorporating  renormalization group requirements.  In Sec. III, we clarify and extend some general properties 
of the RGOPT, that were hinted to before~\cite{rgopt1,rgoptqcd1}, with illustrations in the simpler case of the Gross-Neveu (GN) $O(2N)$ model
where those properties can be compared with exact results as a useful guideline for subsequent QCD considerations. 
We give general arguments why the RGOPT can give reliable nonperturbative approximations
in asymptotically free renormalizable models. In the main section IV we apply the RGOPT to the pion decay constant, 
starting from a well-defined standard perturbative expression,
to extract $F_\pi/\ov\Lam$ values to be compared at three successive RGOPT orders for $n_f=2$ and $n_f=3$.  After examining
in some details the principal problem of generally non-real optimized solutions, we provide a natural and generic cure from appropriate
perturbative renormalization scheme changes. 
In view of realistic determinations  of $\ov\Lam$ and $\alpha_S$
we pay a particular attention to the delicate issue of estimating realistic theoretical uncertainties of the method. 
Section V presents our final results with the extrapolation determining $\alpha_S(\mu)$ values
at different scales with theoretical uncertainties. Finally some conclusions and prospects are given in section VI and the appendix collects
relevant RG and perturbative expressions. 

\section{Renormalization Group Optimized Perturbation}
\subsection{Optimized Perturbation and variations}
The basic feature of the optimized perturbation (OPT) method (which exists in the literature under many names and variations~\cite{delta,kleinert1}), is first to 
introduce an extra {\em unphysical} parameter, $0<\delta<1$ within the relevant Lagrangian, in order to interpolate between ${\cal L}_{free}$ and 
${\cal L}_{interaction}$, in such a way that the mass (here the quark mass 
 $m_q$) becomes a trial or  ``variational'' parameter. In its simplest form: 
\beq
{\cal L}_{QCD}(m_q,\alpha_S) \to  {\cal L}_{QCD}(m (1-\delta), \alpha_S \delta)
\label{LQCDdel}
\eeq
where in our context ${\cal L}_{QCD} (m_q, \alpha_S)$ stands for  
the standard complete QCD Lagrangian,  and  $m_q$ originally is a current quark mass 
relevant for chiral symmetry breaking. 
In the following in practice we shall  
consider basically the cases of two or three (degenerate) light quark flavors $u, d$, or $u, d, s$, thus with corresponding  $SU(2)_L\times SU(2)_R\to SU(2)_V$ 
or  $SU(3)_L\times SU(3)_R\to SU(3)_V$ dynamical
chiral symmetry breaking (incorporating the explicit chiral symmetry breaking from the light quark masses in a later stage, see below).

This procedure is consistent with renormalizability\cite{gn2}, and gauge invariance, provided
that the above redefinition of the QCD coupling $\alpha_S\to \delta \alpha_S$ is performed
consistently for all counterterms and interaction terms
appropriate for renormalizability and gauge invariance. 
At the Lagrangian level it is perturbatively equivalent  
to taking any standard mass-dependent perturbative series in $g\equiv 4\pi\alpha_S$ for a physical quantity $P(m,g)$, {\em after} all 
required mass, coupling etc... renormalizations have been performed,  and to perform the substitution:
\beq m \to  m\:(1- \delta)^a,\;\; g \to  \delta \:g\;.
\label{subst1}
\eeq
Compared with (\ref{LQCDdel}) we introduced an additional parameter $a$, whose role will be explained below, but which can be thought of as 
being fixed at $a=1$ for simplicity at the moment. One then re-expands $P(m,g,\delta)$  
in $\delta$, and takes $\delta \to 1$ afterwards. The exact result in the chiral limit should not depend on the trial mass $m$ artificially introduced, 
but any finite order in $\delta^k$ gives a remnant $m$-dependence. Thus $m$ plays
the role of an arbitrary trial interaction parameter, adjustable order by order. A very often used  and convenient prescription is to fix $m$
at a given order $k$ by optimization (OPT) or ``principle of minimal sensitivity'' (PMS)\cite{pms}: 
\beq
\frac{\partial}{\partial\,m} P^{(k)}(m,g,\delta=1)\vert_{m\equiv \tilde m} \equiv 0\;.
\label{OPT}
\eeq  
One expects optima at successive higher orders to be 
gradually flatter, as indeed confirmed in many applications where higher orders can be worked out~\cite{delta,kleinert1,rgopt1}.\\
Other prescriptions are also possible\cite{delta,kleinert1,pms}, 
like typically looking for plateaus in the $m$-dependence, or imposing ``fastest apparent convergence'' (requiring, at a given order $k$, 
the $k$-th coefficient of the modified perturbative expansion to vanish). But the 
latter two prescriptions often need knowing the original series at relatively high orders to work efficiently. Moreover 
we shall see below that the OPT prescription Eq.~(\ref{OPT}) is particularly well suited when incorporating consistently RG properties, as it simplifies considerably the RG
requirements. \\
Because of the originally massless limit, the interpolation form (\ref{subst1}) is suited to the study of the 
chiral symmetric limit in fermionic models, but it can be easily generalized to scalar field theories 
and to initially massive theories in addition to the trial mass parameter~\cite{delta}. 
In fact the procedure may be viewed as a particular case of ``order-dependent mapping''\cite{odm}, which  
has been proven rigorously to converge\cite{deltaconv,deltac} (exponentially) fast for the energy levels of the $D=1$ $g \phi^4$ anharmonic oscillator model,
exploiting large order and analyticity properties of the oscillator energy.
 The convergence holds even for the double-well oscillator in the strong coupling regime~\cite{deltaconv}, where the standard perturbative series is non Borel summable. 
In very simplified terms the convergence relies on the fact that the 
perturbative expansion for the oscillator energy is a power series in $g/\omega^3$, that makes it possible 
to adjust the trial frequency $\omega$ order by order such as to essentially compensate the factorial growth of the (standard) perturbative 
coefficients at large orders. \\ 
In renormalizable $D > 1$ models, the situation is evidently not so clear, as renormalization gives a more involved $\delta$ series
and mass dependence with logarithmic terms etc... and no rigorous result exists at present on the 
convergence issues of the new series. Although the (linear) $\delta$-expansion (for $a=1$)
was shown~\cite{Bconv} to also damp substantially the generally expected factorial growth of perturbative coefficients at large orders~\cite{gzj_lopt,renormalons},  
it rather appears to delay the ultimate factorial growth. In any case such qualitative large order results are  
of limited practical use to make precise predictions for 
relevant physical quantities, since in many interesting renormalizable models only the very first few
perturbative orders are known exactly.  However, we will give here 
 a simple, both intuitive and RG-consistent argument, supported precisely by exact results in the simpler Gross-Neveu model case,
to understand the empirical stability/convergence OPT properties in renormalizable asymptotically free models.\\

OPT practionners often adopt a more pragmatic attitude, noting that it 
allows to obtain well-defined approximations to nonperturbative quantities beyond the mean field approximations,
which often appear (empirically) quite good at the first few perturbative orders, when results can be confronted  
to other nonperturbative methods, typically lattice simulations. The OPT has also the advantage of being based on ordinary perturbative expressions, 
with well-defined modified Feynman rules and calculations without the eventual complications of other nonperturbative approximations. 
It is thus adaptable to various models~\cite{delta,kleinert1,pms,gn2,qcd1,qcd2,beclde1,beclde2,kleinert2,kastening,bec2,gn3d,optNJL}, 
including the study of phase transitions at finite temperature and density.\\
Now, perhaps a major practical drawback of the method is that the optimization procedure generally introduces
more and more solutions, many being complex, as the perturbative order increases. In cases where there is no 
further insight or constraints from other nonperturbative methods, it may be difficult to choose among the rather 
numerous solutions at higher orders, and the generally complex solutions remain embarassing. \\

Related to this problem, we come now to discuss the additional parameter $a$ introduced in the basic Eq.~(\ref{subst1}): in most applications\cite{delta}, $a$  is set to 1 
and the procedure in Eq.~(\ref{subst1}) is dubbed the linear $\delta$-expansion. However, apart from simplicity and economy of parameters, 
there is no deeper justification for this canonical $a$ value~\footnote{For bosonic models, the linear interpolation is evidently
$m^2\to m^2 (1-\delta)$ instead of Eq.~(\ref{subst1}).}, and other $a$ values may reflect an a priori large freedom 
in the modified interpolating Lagrangian. One can even generalize the interpolation (\ref{subst1}) to introduce several
interpolation parameters~\cite{bec2} at successive orders, depending on the model. The simple optimization prescription
above should be generalized accordingly to fix the extra variational parameters. One may at first naively expect that the procedure will ultimately
converge for different $a$ values, or different interpolating forms, but alternatively  
it is also conceivable that the convergence rate could be optimal only for specific 
values of $a$, depending on a given model. Indeed, in several models, further technical or physical considerations 
do impose specific $a\ne 1$ values beyond the mean field approximation. For the Bose-Einstein condensate (BEC) critical temperature shift, 
evaluated from the OPT approach in different variations~\cite{beclde1,beclde2,kleinert2,kastening,bec2} 
in the framework of the $D=3$ $O(N)$ $\phi^4$ model~\cite{arnold,baymN}, 
the freedom of the interpolating form and extra variational parameters were used~\cite{bec2} to cure the generic problem
of complex optimization solutions, imposing systematically real solutions. 
This reality constraint fixes uniquely $a$ and drastically improves the convergence, with results very close to lattice simulation results.  
Moreover the $a$ value turns out to be related to the anomalous mass dimension critical exponent of the model. At least this 
connection can be identified exactly in the large-$N$ case, and is consistent with results independently obtained by
an alternative interpolating form directly inspired by critical exponent considerations~\cite{kleinert1,kleinert2,kastening}.  
In fact, due to the super-renormalizability of the $D=3$  $\phi^4$ model, the dependence on the (dimensionful) coupling is trivial 
and the relevant perturbative series for the temperature shift, $\Delta T_c \propto \la \phi^2 \ra$, is a 
power series in $g/m$, somewhat like the oscillator energy series.
The idea of modifying the $a$ parameter to impose real optimization solutions has been also used
in the $O(2N)$ GN model case~\cite{rgopt1} for low $N$ values, where at second order in $\delta$ 
it provides results approaching the known exact mass gap below the percent level.
\subsection{RG Optimized Perturbation}
Leaving aside for the moment those reality requirements, while considering more recently the RGOPT  
properties\cite{rgoptqcd1} for the perturbative series relevant for $F_\pi$, a very welcome feature appeared: 
requiring the RG optimized solutions of the (modified) series to be consistent with asymptotic freedom (AF) imposes 
an essentially unique value $a\equiv \gamma_0/(2b_0) \ne 1$, directly related to the anomalous mass dimension (see the appendix for RG
coefficients $\ga_0, b_0$ conventions). This is not a peculiar
feature of the series relevant for $F_\pi$ but, as will be explained below, a more general property of AF models.  
However, unlike the super-renormalizable BEC case,
it does not give at the same time real optimized solutions~\cite{rgoptqcd1}. Reciprocally, if trying
to impose the reality of optimized $g(\ln (\mu/\t m))$ solutions by appropriate $a$ values, following the procedure successful in the BEC case,  
real solutions do not necessarily occur at a given order, or whenever they occur the corresponding
optimized solutions are incompatible with AF. In other words 
the requirements of optimized solutions being {\em both} consistent with
AF and real for a given interpolation (\ref{subst1}) appear mutually incompatible. From general properties and the guidance of
the behavior of exact solutions in the Gross-Neveu model, we will show that the compelling AF
requirement, completely determining the critical $a=\ga_0/(2b_0)$ value, is also crucial for better RGOPT convergence. In a way that 
will be specified below, one can see the OPT modification with critical $a$ as an efficient way to extract a maximal
information even from lowest orders, valid to all orders thanks to RG properties. 
We then propose a more natural way to impose real optimization solutions at arbitrary OPT orders, always compatible with the latter AF properties 
simply by well-defined renormalization scheme changes. 

We now recall the main steps of the RGOPT construction~\cite{rgopt1,rgoptqcd1} for self-containedness.
One considers an ordinary perturbative expansion for a physical quantity $P(m,g)$,
after applying (\ref{subst1}) and expanding in $\delta$ 
at order $k$. In addition to the OPT Eq.~(\ref{OPT}), we require the ($\delta$-modified) series to satisfie a 
standard RG equation:
\beq
  \mu\frac{d}{d\,\mu} \left(P^{(k)}(m,g,\delta=1)\right) =0 
  \label{RGstan}
\eeq 
where the usual homogeneous RG operator
\beq
 \mu\frac{d}{d\,\mu} =
\mu\frac{\partial}{\partial\mu}+\beta(g)\frac{\partial}{\partial g}-\gamma_m(g)\,m
 \frac{\partial}{\partial m} \; 
\label{RG}
\eeq
gives zero to ${\cal O}(g^{k+1})$ when applied to RG-invariant quantities (see Appendix A for our definitions and conventions on RG
functions). Even if the original standard perturbative series to be considered may be (perturbatively) RG-invariant, it is worth noting 
that Eq.~(\ref{RGstan}) provides an independent constraint, not automatically fulfilled, because of 
the non-trivial reshuffling of a part of the perturbative mass to interaction terms, as implied by the interpolation (\ref{subst1}).   
Next note that, combined with Eq.~(\ref{OPT}), the RG equation takes a reduced form:
\beq
\left[\mu\frac{\partial}{\partial\mu}+\beta(g)\frac{\partial}{\partial g}\right]P^{(k)}(m,g,\delta=1)=0
\label{RGred}
\eeq
and Eqs.~(\ref{RGred}), (\ref{OPT}) completely fix (for a given $a$ value in Eq.~(\ref{subst1}))   
optimized values $m\equiv \t m$ and $g\equiv \t g$ since one has two constraints for two parameters.  
A further final simplification occurs, when considering instead of $P^{(k)}(m,g)$ the ratio $P^{(k)}(m,g)/(\ov\Lam)^n$ with $n$
the mass dimension of the operator $P^{(k)}(m,g)$: it is then easy to show that Eq.~(\ref{RGred}) is completely equivalent
to 
\be
 \frac{\partial}{\partial\,g} \left(\frac{P^{(k)}(m,g,\delta=1)}{(\ov\Lam(g))^n} \right) =0 
 \label{RGgopt}
\ee
since by definition $\ov\Lam(g)$ obeys the same equation as (\ref{RGred}) at a given perturbative order. 
Thus we end up with the quite remarkable fact that,
combining the OPT Eq.~(\ref{OPT}) with the RG equation on the dimensionless ratio of the relevant quantity to $\ov\Lam$
amounts to optimizing with respect to the two parameters of the model, $m$ and $g$.~\footnote{In practice using either (\ref{RGred}) or (\ref{RGgopt})
makes no difference, as long as the same RG perturbative order coefficients $b_i$ are used 
consistently in the defining convention for $\ov\Lam$ in (\ref{RGgopt}).}\\

 In summary our RGOPT version involves two important ingredients with respect to most standard OPT studies~\cite{delta}, where essentially 
only the  mass optimization (\ref{OPT}) (or some other prescription on the trial mass) is performed: first, the extra constraint from the
perturbative RG equation (\ref{RGstan}) on the modified series, which in turn, upon requiring perturbative compatibility with asymptotic 
freedom as we will see, will imply a strong restriction on the interpolation form (\ref{subst1}), with a unique critical $a$ value dictated by
the anomalous mass dimension.  
Before proceeding it may be useful at this point to briefly compare qualitatively our present approach with the version 
we investigated years ago~\cite{qcd1,qcd2} to estimate some of the QCD chiral order parameters. 
One major difference  was that, instead of requiring Eq.~(\ref{RGstan}) 
perturbatively as an extra constraint, in \cite{gn2,qcd1,qcd2} we had constructed the 
{\em resummed} pure RG $\delta$-dependence in Eq.~(\ref{subst1}) (but for the linear case with $a=1$) as an integral representation, combined  with 
Pad\'e Approximants (PA) suitably constructed to reach the chiral limit $m\to 0$ without optimizing the mass.
In contrast our new approach only relies on the (modified) purely perturbative information, 
solving the RG and OPT  Eqs.~(\ref{RGred}), (\ref{OPT}) without extra ``knowledgeable input'' beyond perturbative level such as different PA forms. 
This is not only practically more intuitive and simpler to generalize to perturbative expansions for other physical quantities in QCD or other models but, as we 
will examine, the basic interpolation with the critical $a$ value will be much more efficient, exhibiting better empirical stability and convergence properties. 
\section{Guidance from the GN model}
We now make a  digression by considering the main RGOPT features in the case of the GN model in the large-$N$ limit 
to illustrate in this simpler context some remarkable properties of the RGOPT, that are crucial guidelines for the more involved applications in QCD below.  
The case of the mass-gap is essentially a brief reminder of the content of ref.~\cite{rgopt1} while the important case of the vacuum energy was not considered before. 
\subsection{The GN mass gap}
The GN $O(2N)$ model with massless fermions $m=0$
is invariant under the discrete chiral symmetry
$\Psi \to \gamma_5 \Psi $, spontaneously broken so that the fermions get a non-zero mass~\cite{GN}.
At leading $1/N$ order, 
the mass gap is simply 
\be
M_{N\to\infty} = \Lam\equiv  \mu\: e^{-\,\frac{1}{g(\mu)}}\;
\label{MGNinf}
\ee  
in terms of the renormalized coupling~\footnote{The original GN model coupling, defined by $ (1/2) g^2_{GN} (\bar \Psi \Psi)^2 $, 
is convenienly rescaled in the following as $g^2_{GN} N/\pi \equiv g$.} 
$g(\mu)$ in the $\ms$ scheme at the renormalization scale
$\mu$.  The result (\ref{MGNinf}) in the large-$N$ limit can be obtained in different ways by well-known standard calculations. However,
for the time being let us assume  that the only information in the model would come from the purely perturbative expansion of the pole mass, 
in the version incorporating an explicit Lagrangian mass term $m$, and see how the above RGOPT prescription works.
The (renormalized) perturbative expansion of such a mass can be generated to arbitrary perturbative order  
in the large-$N$ case from the compact implicit form~\cite{gn2}: 
\be
M(m,g) = m \left(1+g \ln \frac{M}{\mu} \right)^{-1}
\label{MLNpert}
\ee
where $m\equiv m(\mu)$ and $g\equiv g(\mu)$ are the renormalized mass and coupling
in the $\ms$ scheme. Despite its apparent simplicity, Eq.~(\ref{MLNpert}) generates non-trivial perturbative series at given orders $g^n$, for instance
up to third order:
\bea
\ds && M(m,g)^{(3)} = m \left(1-g L +g^2 (L+L^2) \right. \nn \\
&& \left.  -g^3 (L+\frac{5}{2} L^2+L^3) \right)
\eea
where $L\equiv \ln m/\mu$. From properties of the implicit $M(m)$ defined from (\ref{MLNpert}) and its reciprocal function $m(M)$,  
one can establish~\cite{gn2} that $M(m)\to \Lam$ for $m\to 0$, which provides a consistent bridge between the massive
and massless case. But this needs the knowledge of the all order expression (\ref{MLNpert}), only known exactly in the large-$N$ limit.\\
Now alternatively, performing substitution (\ref{subst1}) (fixing $a=1$ at the moment) on (\ref{MLNpert}), expanding at
a given $\delta^k$-order and taking $\delta\to 1$, (\ref{RGred})
has the non-trivial solution~\cite{rgopt1}
\be
\t g= -1/L \equiv (\ln\frac{\mu}{m})^{-1}
\label{gLN}
\ee
already at first order.
At arbitrarily high perturbative order $k$ the RG equation factorizes to a form
such that $g=-(L)^{-1}$ is always a solution.
(\ref{gLN}) reminds of the perturbative expression of the running coupling $g$ for $\mu\gg m$
(the exact running coupling for $N\to \infty$ being given by Eq. (\ref{gLN}) but with $m\to \Lam$).
Moreover, injecting this RG perturbative behavior solution directly into $M^{(k)}(m,g,\delta=1)$ simply 
gives at arbitrary order $k$ 
\be
M^{(k)}(m,g,\delta=1)= m
\label{M=m}
\ee
without any extra correction, having not yet used at this stage the OPT equation determining
$m$. 
Then  using (\ref{gLN}) the OPT equation (\ref{OPT}) gives at arbitrary order $k$:
\be 
\t L \equiv \ln \frac{\t m}{\mu} = -1\; .
\label{optLN}
\ee
Thus the exact mass gap is obtained already at the very first RGOPT order, as well as 
$\t m =\Lam$, which together with (\ref{gLN}) also gives the known exact running coupling. \\
However, starting at third order extra spurious 
solutions appear in either OPT or RG equations: for example the RG
equation has an extra solution: 
\be
g = -(2+5L+2L^2)^{-1}
\label{RGN3}
\ee     
having clearly the wrong RG-behavior for large $L$, and leading to a very odd
result for the mass gap\cite{rgopt1}: $M/\Lam = -{\rm e}^{-(25/2)}/324$. At even higher orders all kinds of spurious solutions appear,
most being complex not surprisingly.\\ 

The important remark is that all those spurious solutions at higher order can be easily rejected,  
even if not knowing the correct exact result, on the
basis that they do not have the correct perturbative RG
behavior, with the right coefficient for AF.
This compelling requirement can be directly translated into a similar one for the QCD case~\cite{rgoptqcd1}, as we will detail below.\\
This exact coincidence between the mass-gap $M$ and the (originally perturbative) 
mass $\t m$ after OPT is performed, meaning that there are no further perturbative corrections, is certainly 
a peculiar feature of the large-$N$ limit, but nevertheless a non-trivial direct consequence 
of the RGOPT construction, not obvious from the original perturbative expansion of Eq.~(\ref{MLNpert}).
Indeed truncating at arbitrary finite order the standard perturbative series (\ref{MLNpert}), one only finds the trivial result $M(m,g)\to 0$ for $m\to 0$, 
while the correct result can only be obtained by having the possibility, in this simple case,  
of extracting the $m\to 0$ limit from the properties of the {\em all order} series defined from (\ref{MLNpert}). In that sense the RGOPT appears to 
perform an efficient shortcut, using a minimal amount of information from perturbation to obtain the correct result already at first order. \\
One may suspect at first sight that the above properties are only a consequence of the peculiar mathematical properties of expression (\ref{MLNpert}), related to the relatively
simple large-$N$ properties of the GN model. 
For arbitrary $N$ values, where the equivalent of (\ref{MLNpert}) is only known perturbatively to lowest orders, instead of the simple result in (\ref{M=m})
one obtains~\cite{rgopt1} for the optimized mass after RGOPT $\t m = c(N) \Lam$ with $c(N)$ a constant relatively close to 1 (the closer as $N$ increases). 
We will see that most of those features survive analogously, approximately, in a more general AF model like QCD.\\

The extra parameter $a$ in (\ref{subst1}) was fixed to $a=1$ in these considerations. If one takes other values of $a$,
the property of getting the exact large-$N$ mass gap at any order is lost, as it appears that any value $a \ne 1$ is not compatible with the exact RG solution  (\ref{gLN}).
Forcing nevertheless $a \ne 1$, the behavior of solutions is more obscure but empirically seems to still converge,  much more slowly.  
So for the GN mass gap in the large-$N$ limit, taking $a=1$ gives the fastest optimal convergence rate.
 Before inferring general statements
from this very particular case, however, we shall see next the case of the vacuum energy which is a little more subtle. 
\subsection{The GN vacuum energy}
The GN model vacuum energy $E_{GN}$ can be evaluated in the chiral symmetric and large-$N$ limit, and reads~~\cite{EGNorig}
\be
E_{GN} = -(\frac{N}{4\pi}) \Lam^2\;.
\label{vacex}
\ee
Alternatively its perturbative expansion can be expressed, after all necessary renormalizations, in the
following compact form~\cite{gn2} in terms of the explicit mass $m$ and the mass gap $M(m,g)$ defined in Eq.~(\ref{MLNpert}):
\be
E_{GN} = -(\frac{N}{4\pi})\,\left( M^2(m,g) +2\frac{m}{g} M(m,g)\right).
\label{vacGN}
\ee
Again the purpose here is to examine what can be obtained from the  RGOPT when starting from a purely perturbative information, truncating
(\ref{MLNpert}) and (\ref{vacGN}) at a limited order $g$. 
Similarly to the mass gap case above, the remarkable feature is that after performing (\ref{subst1}) expanded to any arbitrary order $\delta^k$,
setting $\delta=1$ etc, Eq.~(\ref{RGred}) always gives the exact solution (\ref{gLN}), leading also to the OPT solution (\ref{optLN}) and to the
exact result (\ref{vacex}), together with $m =\Lam$. Also similarly, using solely Eq.~(\ref{RGred}) and its solution (\ref{gLN}) within
the $\delta$-modified perturbative series of the vacuum energy, at any orders it already entails $E_{GN} =-N/(4\pi)\, m^2 $ without any perturbative extra corrections.\\ 
However all these results are obtained when taking $a=1$. For arbitrary $a$, 
the RG equation (\ref{RGred}) at orders $\delta^k$ happens to give a solution $g(\ln m/\mu)$ only compatible with AF
for integer and half-integer $a \ge 1$ values, the larger values appearing successively at increasing orders:
\be
RG(g=-L^{-1}) \propto \Pi_{i=0}^k (a-1-\frac{i}{2})\equiv 0\;.
\label{a_gn}
\ee
{\it i.e.} Eq.~(\ref{RGred}) is only compatible with $g=-L^{-1}$ for $a=1$ or $a=3/2$ at order $\delta$; for $a=1, 3/2, 2$ at order $\delta^2$, and so on.\\
The result (\ref{a_gn}) is in fact the particular $N\to\infty$ limit for the GN model of a more general one, first noticed for the QCD perturbative 
series relevant for $F_\pi$~\cite{rgoptqcd1}:
as will be explained in more details below, the RG equation at order $\delta^k$ 
gives a solution $g(L)$ compatible for $g\to 0$ with AF,  
\be
\t g (\mu \gg \t m) \sim (-2b_0 L)^{-1}+{\cal O}(L^{-2})\;,
\label{rgasympt}
\ee
{\em only} if $a$ takes specific discrete values, appearing at successive orders:
\be
RG \propto \Pi_{i=0}^k (a-\frac{\gamma_0}{2b_0}-\frac{i}{2}) \equiv 0\;.
\label{RGcrit}
\ee
In the GN model $\gamma_0/(2b_0)\equiv (N-1/2)/(N-1)\to 1$ in the large-$N$ limit.\\ 
However all other values $a\ne \ga_0/(2b_0)$ appearing at higher orders lead to some pathological behavior: either 
the OPT equation (\ref{OPT}) cannot be satisfied (while it is always compatible for $a=\ga_0/(2b_0)$), or when both RG and OPT equations are compatible, the final
result is wrong by a large amount in the GN model case. For example, at order $\delta^2$, for $a=3/2$, $g=-L^{-1}=1/2$ is a unique solution, but it gives $E_{GN}=1/2 E_{GN}(\rm exact)$.
Similarly, for $a=2$, still at order $\delta^2$, $g=-L^{-1}=-1/2$ is unique solution, but gives $E_{GN}= -E_{GN}(\rm exact)$. Moreover, even if relaxing the
AF constraint $g=-L^{-1}$, one can then find solutions at successive orders for those other $a\ne1$ values, but most are complex, and though some solutions
have small imaginary parts and seem to converge very slowly to the correct vacuum energy result, their general trend is not conclusive.
Hence, this is a strong indication that the value $a=\gamma_0/2b_0$, the only one valid for all $\delta^k$ orders, 
should give the best convergence rate. 
\subsection{General properties and guidance for QCD}
In a more involved theory like QCD it seems at first more difficult to guess a right optimized RG solution among many appearing at higher and higher orders
from a blind optimization. But a crucial guidance is the rejection 
of most {\it a priori} spurious optimized solutions if 
requiring~\cite{rgoptqcd1} the RG solution of (\ref{RGstan}) to have 
the correct perturbative RG behavior compatible with AF (\ref{rgasympt}), which is only possible   
for a critical value of $a$ in the interpolation (\ref{subst1}):
\be
a =\frac{\gamma_0}{2b_0} =\frac{12}{33-2n_f}
\label{acrit}
\ee
where the relevant value in QCD is $a=12/29\, (4/9)$ for $n_f=2$ (respectively $n_f=3$). 
The maximally fast convergence for (\ref{acrit}) can be inferred from a general argument as we examine now. \\

Coming back to the vacuum energy expression (\ref{vacGN}), one notices 
that after performing (\ref{subst1}) with $a=1$ and using the optimized solutions, 
the second term, $\propto m$, vanishes identically at any order, thus giving
no contribution to the optimized energy. This GN result is in fact a special case of a more general one. Let us consider~\cite{gn2,qcd2,Bconv} 
\be
M(m,g)\equiv m [1+2b_0 g \ln \frac{M}{\mu}]^{-\frac{\gamma_0}{2b_0}}\equiv \hat m\,[\ln \frac{M}{\Lam_0}]^{-\frac{\gamma_0}{2b_0}}
\label{MRG0}
\ee
defining $M(m,g)$ implicitly, where we introduced the standard scale invariant mass $\hat m \equiv m (2b_0 g)^{-\ga_0/(2b_0)}$ and
scale $\Lam_0\equiv \mu e^{-1/(2b_0\,g)]}$ consistent at first RG order to make the RG-invariance of (\ref{MRG0}) explicit.
This is the straightforward generalization of (\ref{MLNpert}) in an AF model with arbitrary first order RG 
coefficients $b_0$ and $\gamma_0$, and which correctly resums the $\ln m$ dependence to all orders.  
Now consider the perturbative expansion for RG-invariant quantities formally written as 
\be
R^{p,q}(m,g) =(\frac{\hat m}{M})^{p \frac{2b_0}{\ga_0}} M^q = (\ln \frac{M}{\Lam_0})^p  M^q\;,
\label{Mgen}
\ee
where  from (\ref{MRG0}) the two forms are completely equivalent, and generalize for $\ga_0/(2b_0)\ne 1$ 
the two terms in (\ref{vacGN}) with $q=2$ and $p=0, 1$ respectively. Then  performing the OPT on expressions (\ref{Mgen}) with an arbitrary 
$a$ in (\ref{subst1}), it can be shown after some algebra that  
\begin{itemize}
 \item i) for $p=1$, requiring AF compatibility (\ref{rgasympt}), the non-vanishing part of the RG equation for $g\to 0$ at 
 arbitrary $\delta^k$ orders takes a factorized form similar to (\ref{RGcrit}), thus necessarily requiring corresponding 
 critical $a$ values. For $p=0$, the RG equation alone appears AF compatible independently of $a$ values, but
compatibility with the OPT equation also necessarily 
requires uniquely $a=\ga_0/(2b_0)$. Note that the constraints on $a$ do not involve higher order RG coefficients 
$\ga_i$, $b_i$ for $i\ge 1$ appearing at 
higher orders: the latter terms enter subleading logarithms, while the previous properties, as far as concerns the leading AF behavior (\ref{rgasympt}), 
are fully determined by the leading logarithm terms, which at any order only depend on
$b_0$ and $\ga_0$. Consequently $a$ remains solely determined by the
first RG order coefficients. 
 \item ii)  Taking thus $a=\ga_0/(2b_0)$, the RG Eq.~(\ref{RGred}) applied to $R^{p,q}(m(1-\delta)^a,\delta g,\delta\to 1)$ in Eq.~(\ref{Mgen}) 
 at arbitrary orders $\delta^k$, takes the factorized form $\mbox{RG}(g,L)=(1+2b_0 L g) f(g,L)$ thus with a unique exact solution
 \be
\t g \t L =-\frac{1}{2b_0}
\label{rg0gen}
\ee
plus other spurious solutions not consistent 
with AF for small $g$ (large $|L|$).
\item iii) combining the RG solution (\ref{rg0gen}) with the OPT Eq.~(\ref{OPT}) gives a single ($k$-degenerate) solution 
\be
\t L=-\frac{\ga_0}{2b_0}\;;\;\; \t g=\ga_0^{-1}\;.
\label{opt0gen}
\ee
\item iv) Using  (\ref{rg0gen}) alone, substituting indifferently for $\t g$ or $\t L$, (\ref{Mgen})
gives simply $(\Lam_0)^q$ for $p=0$,  while it vanishes for $p=1$, at any $\delta^k$ order.
\end{itemize}
Moreover, very similarly to the GN case, values of $a=\gamma_0/2b_0+k/2$ for integers $k \ne 0$, are generally not 
compatible with the OPT Eq.~(\ref{OPT}) or give also rather pathological or very unstable behaviors. 
Even in the absence of pathological behavior, since we have at our disposal only a few successive perturbative terms, it makes sense anyway 
to follow and compare successive orders with the value $a=\gamma_0/2b_0$ valid for any $\delta^k$ orders. 
In contrast any other $a$ value does not lead to the simple and exact properties above in i)-iv), rather giving relatively unstable
optimized solutions at successive orders with no obvious pattern and convergent behavior. \\

 The RG-invariant quantities in (\ref{Mgen}) may appear somewhat formal, but in the approximate world 
where only the first RG order would contribute, the relevant perturbative 
expansion  at arbitrary orders of a physical quantity of mass dimension $q$ would be a linear combination of the two simple RG-resummed forms in (\ref{Mgen}), 
for $p=0,1$. We will see below a concrete example for the 
actual perturbative series for $F_\pi$. The above results thus show that the RGOPT performed with  (\ref{acrit})  
would immediately select, already at first order, the relevant ``nonperturbative'' pieces $\propto \Lam_0$ exactly while 
discarding the ``spurious'' purely perturbative terms that do not survive the $m\to 0$ limit. 
In most models the complete perturbative series take evidently a more involved form when including higher orders, not only 
the higher order RG dependence no longer takes the closed form (\ref{MRG0}) but also 
the non-RG perturbative contributions at successive orders should be included. Nevertheless,  
since a series for a physical quantity is perturbatively RG-invariant, it is always possible in principle 
to re-express it as linear combinations of explicitly RG-invariant forms, appropriately generalizing (\ref{MRG0}) at higher orders. 
For instance this can be done explicitly~\cite{qcd2,Bconv} for the exact two-loop RG dependence to all orders in a relatively compact 
form. In fact, a consistent (RG invariant) perturbative series will automatically build order by order the correct 
logarithmic and non-logarithmic coefficients that would be dictated from such explicitly RG-invariant resummation like (\ref{MRG0}), so
the complete resummed expressions are not even needed to perform the RGOPT at low perturbative orders, often the only ones available in practice. \\ 

But for such complete perturbative series incorporating non-RG and RG terms beyond first order, applying the RGOPT with (\ref{acrit}) 
no longer gives the simple results (\ref{rg0gen}), (\ref{opt0gen}), not surprisingly. First, even the pure RG dependence should  
involve the two-loop RG coefficients $b_1, \ga_1$. More importantly the non-logarithmic contributions at each perturbative orders imply anyway a 
departure from this pure RG behavior. 
Nevertheless, the properties in i) above, completely determining $a=\ga_0/(2b_0)$ from AF compatibility, remain exact, and at arbitrary orders, since these
only depend on the leading logarithm behavior as explained above. Moreover, although the other exact results ii)-iv) are lost,  
some properties of the above simple picture  remain approximately when performing ``blindly'' the OPT for an 
arbitrary series with the prescription (\ref{acrit}). More precisely,
for a physical quantity of mass dimension $q$, defining formally its perturbative series:
\be
P^{(q)}(m,g) = m^q f(g,\ln \frac{m}{\mu})\;,
\label{PQ}
\ee 
the RGOPT transmutes~\footnote{After 
having performed the RG or OPT (\ref{RGred}) or (\ref{OPT}), solved for the mass, the resulting series may be intuitively viewed as coming from  perturbative Feynman graphs
with original masses replaced by dressed masses of order  $\ov\Lam$, having a nonperturbative $g$-dependence. But an explicit modified graph picture
is not necessary for any practical computation.} this series, which had a trivial
chiral limit $m\to 0$, into a series where mass and coupling optimization give  
$\t m \simeq c \ov\Lam$  with $c$ of order 1 depending on the details of the model, the RG and non-RG perturbative coefficients.
Coming back to the original motivation for the OPT, this is quite satisfactory: one expects that the physical result 
after modifying the series should depend as little as possible on the artificially introduce mass $m$, and indeed, 
the optimized mass is fully determined by the only scale $\ov\Lam$ in the theory. 
Now in the present case {\em if} in addition  
the optimized coupling $\t g$ remains reasonably perturbative (which is correlated with $\t m$ remaining of order $ \ov\Lam$), 
the optimized result cannot depart too much from its simple one-loop form  $g=-1/(2b_0 \t L)$ above.
The departure is entirely determined by the optimization and the remaining ``perturbative'' corrections to the bulk result
$P^{(q)} \simeq \t m^q= (c \ov\Lam)^q$, entering in $f(\t g, \ln \t m/\mu)$ in Eq.(\ref{PQ}),  
are now expected to be a moderate correction. \\

Since the above mechanism is quite generic, we expect the same feature to occur in
any renormalizable AF model, whatever the details of the RG and other coefficients.
This is precisely what will happen for the series relevant to $F_\pi$, as we will explore in details at three successive orders, where we will
see that even for the complete perturbative series the optimized solutions do not depart much from the relation (\ref{rg0gen}), with moreover
a stabilization of the optimized coupling and mass towards more perturbative values as the order increases. 
These features give an intuitive key argument for the empirically seen stability/apparent convergence of the method to which we now come.
\section{The pion decay constant}
We now arrive at our main application of the RGOPT on a perturbative series that is 
directly connected to the pion decay constant $F_\pi$. 
One very convenient definition of $F_\pi$ 
is via its connection to the axial current-axial current correlator, that is very familiar {\em e.g.} in the context of Chiral Perturbation
Theory (ChPT) where it appears as the lowest ${\cal O}(p^2)$ term~\cite{chpt}. More precisely:
\beq
i \langle 0| T A^i_\mu (p) A^j_\nu (0)|0 \rangle \equiv \delta^{ij}
 F^2_\pi g_{\mu\nu} +{\cal O}(p_\mu p_\nu) \label{Fpidef}
\eeq
where  the axial current is $A^i_\mu \equiv \bar q \gamma_\mu \gamma_5 \frac{\tau_i}{2} \:q$ for $SU(2)$
(or for $SU(3)$, $\tau_i\to\lambda_i$ where $\lambda_i$ are the Gell-Mann matrices),
and in this normalization $F_\pi\sim 92.3$ MeV~\cite{PDG}. Actually, to be more precise, because of the chiral limit implied by the OPT
method, we should consider that after OPT we will consistently obtain from (\ref{Fpidef}) $F_\pi$ in the strict chiral limit, 
{\it e.g.} for $SU(2)$ $F\equiv F_\pi(m_q\to 0)\simeq 86$ MeV~\cite{chpt}. We will consider below recent determinations of $F_\pi/F$ (or similarly $F_\pi/F_0$ in the
$SU(3)$ case) to be properly taken into account in our analysis.
Nevertheless we need to start from a perturbative expression in terms of explicit quark masses in order to apply the RGOPT. 
The obvious advantage of using the above expression in our context 
is that the standard QCD perturbative expression of the correlator in Eq.(\ref{Fpidef}) in the $\ms$ scheme,  is presently fully 
known analytically up to four-loop ($\alpha^3_S$) contributions~\cite{Fpi_3loop,drho3loop,Fpi_4loop}. 
This welcome feature will allow us to compare the results
of our approach at three successive orders and thus provide some definite outcome on the stability/convergence properties, in a full QCD framework.
\begin{figure}[htb]
\epsfig{figure=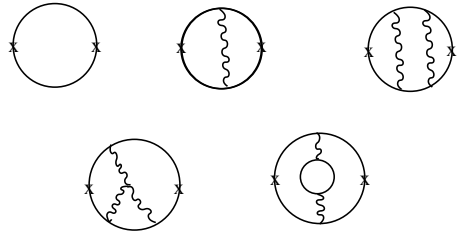,width=6cm}
\caption[long]{Samples of standard perturbative QCD contributions to the two-point axial-vector correlator up to three-loops. 
Crosses denote axial current insertions.}
\label{fpipert}
\end{figure}
With a slightly adapted change of normalization,
it reads in our notations: 
\bea 
& F^2_\pi(\rm pert) = 3 \frac{m^2}{2\pi^2} \left[ \mbox{div}(\epsilon,\alpha_S) -L +\frac{\alpha_S}{4\pi}(8 L^2
+\frac{4}{3} L +\frac{1}{6}) \right. \nonumber \\ 
&\left. +(\frac{\alpha_S}{4\pi})^2 (f_{30} L^3 +f_{31} L^2 +f_{32} L +f_{33})  \right.\nonumber \\
&\left. +(\frac{\alpha_S}{4\pi})^3 (f_{40} L^4+f_{41} L^3 +f_{42} L^2+f_{43} L +f_{44})
\right]  \label{Fpipert}
\eea    
where $m$ is the {\em running} mass in the $\ms$ scheme, $L\equiv \ln \frac{m}{\mu}$, 
and the three-loop $f_{3i}$ and four-loop $f_{4i}$ coefficient expressions,  
known for an arbitrary number of flavors $n_f$, are given 
in the appendix. A few representative Feynman graphs contributions at successive orders up to three-loops are illustrated in Fig.~\ref{fpipert} (there
are evidently many more contributions not shown here). Note that the one-loop order is ${\cal O}(1)={\cal O}(g^0)$ and does not depend on $\alpha_S$. 
As a rather technical remark, note that the originally calculated expressions in refs.~\cite{Fpi_3loop,drho3loop,Fpi_4loop} are
generally known for $n_h$  massive quarks and $n_l$ {\em massless} quarks entering at three-loop order, but are 
often given for the specific case $n_l= n_f-1, n_h=1$ more relevant in various QCD applications. However, in our context $m$ is the ($n_f$-degenerate) light quark mass and its precise mass dependence
is what is relevant for the optimization procedure, so one should trace properly the full $n_l, n_h$ dependence,  
to take then $n_l=0$ and $n_h\equiv n_f$ with $n_f=2 (3)$ for the $SU(2)$ (resp. $SU(3)$) 
case.\footnote{Some results in \cite{Fpi_3loop} partly in the on-shell scheme, need to be converted in the $\ms$-scheme, using the 
two-loop~\cite{Mpole2l} and three-loop~\cite{Mpole3l} pole-to-running mass relations. The coefficients $f_{ij}$ in the $\ms$-scheme 
of all the logarithmic terms $L^k, k\ge 1$ were deduced consistently up to four-loop $\alpha^3_S$ order from lowest orders using also 
RG properties and explicitly crosschecked with direct calculations~\cite{PrivComm}. At three-loop (and four-loop) order one should also take care 
to extract the non-singlet contributions to the axial vector correlator, only relevant to (\ref{Fpidef}).}. Now, anticipating 
on the results below, we note that the optimization results are actually not very sensitive to the detailed 
$n_f$-dependence entering at three-loop and higher orders: as we will examine precisely, even an arbitrary change of a factor $\sim 2$ typically 
in the three-loop coefficient $f_{33}(n_f)$ induces a change of 3-4\% at most in the final optimized $F/\ov\Lam$ results.
\subsection{Renormalization and RG-invariance}
There is however one subtlety at this stage: the calculation in dimensional regularization
of (\ref{Fpidef}) actually still contains divergent terms needing extra subtraction after mass and coupling renormalizations in 
$\ms$ scheme, formally indicated as $\mbox{div}(\epsilon,\alpha_S)$ in Eq.~(\ref{Fpipert}), whose explicit expression is given in the appendix. 
This is most simply seen already at first order, where the only one-loop contribution in Fig.~\ref{fpipert}, proportional 
to $m^{2-\epsilon}_0/\epsilon$ in dimensional regularization with $D\equiv 4-\epsilon$,  is independent of  $g$ and this divergence 
cannot be removed by the mass renormalization affecting only the next order, $m_0\to m Z_m = m (1+{\cal O}(g/\epsilon))$. 
The correct procedure to obtain a RG-invariant finite expression while subtracting those divergences consistently is well-known from 
standard renormalization of composite operators with mixing~\cite{Collins}.
(Although to our knowledge it has been seldomly applied to the present series, as in most relevant applications (\ref{Fpipert})
only enters in combinations where those divergences cancel out, like typically in the electroweak $\rho$-parameter calculations~\cite{drho,drho3loop}, where 
up to an overall constant (\ref{Fpipert}) is  related to the (non-singlet) contributions to the $Z$-boson self-energy, with $m\simeq m_{top}$). \\  
We can define~\cite{gn2,qcd2} the needed subtraction as a perturbative series:
\be
{\rm sub}(g,m) = \frac{m^2}{g} H(g) \equiv \frac{m^2}{g} \sum_{i\ge 0} s_i  g^{i}
\label{sub}
\ee
with coefficients determined order by order by
\be
\mu \frac{d}{d\mu}{\rm sub}(g,m) \equiv {\rm Remnant}(g,m)=\mu \frac{d}{d\mu}[F^2_\pi({\rm pert})|_{\rm finite}]
\label{rganom}
\ee 
where the remnant part is obtained by applying the RG operator Eq. (\ref{RG}) to the 
finite part of (\ref{Fpipert}), not separately RG-invariant. 
Thus the (finite) quantity:
\be
F^2_\pi({\rm pert})|_{\rm finite}-{\rm sub}(g,m) \equiv F^2_\pi({\rm pert})|_{\rm RG inv.}
\label{FpiRG}
\ee
is by construction RG-invariant
at a given order. Note that (\ref{sub}) does not contain any $\ln m/\mu$ terms and necessarily 
starts with a $s_0/g$ term to be consistent with RG invariance properties, as the one-loop contribution in Eq.~(\ref{Fpipert})
is of order $1$. To obtain RG-invariance at order $g^k$, fixing $s_k$ in (\ref{sub}), one needs knowledge of the coefficient of the $L$ term 
(equivalently the coefficient of $1/\epsilon$ in dimensional regularization) at order  $g^{k+1}$.
 This renormalization procedure is completely similar
for the GN model vacuum energy above discussed, where the required subtractions contribute to the second term in Eq.~(\ref{vacGN}). 
A familiar completely analogous 
case is the so-called anomalous dimension of the QCD vacuum energy,  entering in the renormalization
procedure of the $m \qq$ operator due to mixing~\cite{vac_anom} with $m^4 1$. 
It can be derived consistently by following the very same procedure as above. 
The $s_i$ coefficients can be expressed in terms of RG coefficients and other terms using RG properties. In more compact form they read:
\bea
&& s_0 = \frac{3}{4\pi^2(b_0-\gamma_0)}\;\;=\frac{4}{1-\frac{2}{9}\,n_f}\;, \nn \\
&& s_1= 3\left(-\frac{5}{24\pi^2} +\frac{1}{2} \frac{\ga_1-b_1}{\ga_0-b_0}\right)= \frac{1}{16\pi^2} \frac{(237 + 17n_f)}{(-9 + 2n_f)}, \nn \\
&& s_2 =  \frac{ (-78777 + 369 n_f + 619 n^2_f + 432 (9 + 38 n_f)\zeta(3))}
{384\pi^4 (-57 + 2 n_f)(-9 + 2 n_f)},\nn \\
&&  s_3\simeq  \frac{3} {(-57 + 2 n_f) (-45 + 2 n_f) (-9 + 2 n_f)} \times \nn \\
&&(1.2569 - 0.6799 n_f + 0.0451 n^2_f -8.5\,10^{-4} n^3_f  - 3.1\,10^{-5} n^4_f)\;.\nn \\
\label{s_i}
\eea
Equivalent results are obtained more formally 
by working with bare expressions, establishing the required RG properties between bare and renormalized quantities. The subtraction
$H(g)$ in Eq.~(\ref{sub}) is then entirely determined by the coefficient of the simple $1/\eps$ pole in (\ref{Fpipapp}) (see the appendix for more details). 
\subsection{RG optimization at successive $\ms$ orders}
We are now ready to  apply to Eq.~(\ref{Fpipert})-(\ref{sub}) the procedure (\ref{subst1}) 
and expand at order $\delta^k$, then solve OPT and RG Eqs.(\ref{OPT}), 
(\ref{RGred}) (or equivalently (\ref{RGgopt})). 
Before embarking into higher orders, it is worth working out the RGOPT successive
steps at the simplest non-trivial $\delta^1$ order, where everything is very transparent. Furthermore, we shall consider
the approximation where solely the first order RG dependence $b_0$, $\ga_0$ is taken into account, 
suppressing thus also non-logarithmic contributions. This may be considered a crude approximation but it will in fact nicely illustrate
the remarkable properties discussed in Sec. III C. Extracting thus from (\ref{Fpipert}) together with the subtraction (\ref{sub}) the
terms depending only on  $b_0$, $\ga_0$ up to (two-loop) order $g$ one finds the following expression:
\bea 
&& F^2_\pi(\mbox{RG-1},{\cal O}(g)) = 3 \frac{m^2}{2\pi^2} \left[ -L +\frac{\alpha_S}{4\pi}(8 L^2
+\frac{4}{3} L) \right. \nn \\
&&\left. -(\frac{1}{8\pi(b_0-\ga_0)\,\alpha_S} -\frac{5}{12})\right]
\label{Fpi1}
\eea
where the last two terms correspond to the subtraction with coefficient $s_0, s_1$ in (\ref{s_i}) (but $s_1(\mbox{RG-1l})=-5/(8\pi^2)$ being different now 
due to the approximation $b_1=\ga_1=0$, see Eq.~(\ref{s_i})).
Performing (\ref{subst1}) with (\ref{acrit}) at order  $\delta$, for example for $n_f=2$,  
straighforward algebra gives the modified series
\bea
&& F^2_\pi(\mbox{RG-1},{\cal O}(\delta))|_{\delta\to 1} =  3 \frac{m^2}{2\pi^2} \nn \\
&& \left[-\frac{102\pi}{841\,\alpha_S}+\frac{169}{348}-\frac{5}{29}L+\frac{\alpha_S}{4\pi}
(8 L^2 +\frac{4}{3} L) \right]
\label{Fpid1}
\eea
In passing one sees here a general property of OPT, that at a given $\delta^k$ order only lower orders $g^p, p <k$ are modified by (\ref{subst1}), while 
the order $g^k$ (the order $\alpha_S$ in (\ref{Fpi1})) remained untouched. 
The RG equation (\ref{RGred}), consistently limited at first order $b_0$, gives two solutions, one being (\ref{rg0gen}) exactly. 
The OPT equation alone has more complicated solutions $L(g)$, but combined with (\ref{rg0gen}) simply gives the unique solution
(\ref{opt0gen}):
\be
\t L=-\frac{\ga_0}{2b_0}\;;\;\;\t \alpha_S =\frac{\pi}{2}\;,
\label{pureRG1}
\ee
fully confirming the general properties mentioned in Sec.III after Eq.~(\ref{Mgen}) for the concrete $F_\pi$  perturbative series. 
Finally, substituting (\ref{rg0gen}) either viewed as $\t g(L)$ or $\t L(g)$ within (\ref{Fpid1}) gives
\be
F_\pi(\mbox{RG-1},{\cal O}(\delta))(\t m,\t g)= (\frac{5}{8\pi^2})^{1/2} \t m \simeq 0.25  \Lambda_0
\label{Fpid1fin}
\ee
where we identified the basic scale $\t m \equiv \Lambda_0= \mu e^{-1/(2b_0 g)}$ at this first RG order. Actually the final ratio $F_\pi/\Lambda_0$ in 
(\ref{Fpid1fin}) is independent of $n_f$, a peculiarity of this pure first RG order approximation (the $n_f$ dependence enters only via $b_0(n_f)$ in $\Lambda_0$). 
We also see here the announced property that the optimized mass is of order $\ov\Lam$, in fact exactly in this approximation. 
Numerically this is already a quite realistic value of $F_\pi/\ov\Lam$, though evidently at higher orders 
$\ov\Lam$ should include higher RG orders.  
Note also that Eq.~(\ref{opt0gen}) does not play any role, again a peculiarity of this pure RG approximation. 
The optimized coupling $\t \alpha_S =1/(4\pi \ga_0) =\pi/2$ is relatively large for QCD, but it does not matter in this approximation where 
there are no further perturbative corrections to the relation (\ref{Fpid1fin}). 
 It is instructive to see this result in terms of the more formal explicitly RG-invariant forms introduced above in (\ref{Mgen}). It is easily shown
that the complete terms in the original series (\ref{Fpi1}) can be obtained from the perturbative re-expansion to ${\cal O}(g)$ of two such invariants:
\be
F^2_\pi(\mbox{RG-1},{\cal O}(g)) = \left[ -\frac{s_0}{2b_0} R^{1,2}(m,g) -s_1 R^{0,2}(m,g)\right]_{{\cal O}(g)}\;.
\ee
Thus, using property iv) after Eq.~(\ref{Mgen}), one could have derived directly that performing the RGOPT at any order (in this approximation),
only the last term survives, to give $F^2_\pi(\mbox{RG-1}) =-s_1\, \Lam_0^2$ (where $b_1=\ga_1=0$ in $s_1$ in (\ref{si_exact}), in agreement with (\ref{Fpid1fin})).\\
In fact the calculation based on pure RG dependence can also be done exactly up to the next RG order, {\it i.e.} including the two-loop RG coefficients
$b_1, \ga_1$ consistently (but still neglecting the non-logarithmic terms) to order $g$ in (\ref{Fpipert}). The RG equation (\ref{RGred}) 
still has an exact relatively simple solution, generalizing (\ref{pureRG1}): 
\be
-2b_0 \t L \t g =1\;;\;\;{\t g}^{-1} =(4\pi \t \alpha_S)^{-1} =\ga_0 -\frac{\ga_1-b_1}{\ga_0-b_0}\;.
\label{pureRG2}
\ee
with the first simple relation between the optimized coupling and the mass still valid. 
Putting numbers, (\ref{pureRG2}) gives: $\t\alpha_S = 6\pi (9-2n_f)/(255+13n_f) \simeq 0.33(0.19)$; $\t L\simeq -1.94(-3.63)$
respectively for $n_f=2 (3)$. Thus including the pure two-loop RG order dependence 
gives a drastic reduction of the optimized coupling and related mass, having more perturbative values.  
In contrast with the first RG order, however, the OPT Eq.~(\ref{OPT}) cannot be fulfilled exactly for (\ref{pureRG2}), but
gives a small remnant term, proportional to $(\ga_1-b_1)/b_0$ when normalized to one, thus
formally of higher order $\sim 1/(16\pi^2)$. Plugging the solution (\ref{pureRG2}) within the corresponding $F_\pi$ expression at order $\delta$ 
gives $F/\ov\Lam\simeq 0.2 (0.18)$ respectively for $n_f=2(3)$. \\ 
At higher orders for a fully realistic determination
the OPT will consistently incorporate RG (and non-RG) higher order dependence specific to the $F_\pi$ series available from (\ref{Fpipert}), 
and Eq.~(\ref{OPT}) will now play a non-trivial role, producing a departure of the optimized $\t g$ (or $\t L$ equivalently) from these pure RG values. 
Such properties are somewhat hidden in the calculational  
details but can be fully controlled a posteriori (since exactly tractable solutions of polynomial equations are involved), 
with the results that the induced departure from the pure RG results will be a reasonable perturbation. 
 Indeed as we will see next, the true optimized solutions for $F_\pi$ at higher orders lie somewhere in   
between the pure RG first order (\ref{pureRG1}) and second order (\ref{pureRG2}) results, with decreasing and finally stabilizing optimized coupling at increasing
orders. This self-adjustement of the coupling $\t g$ with the order is indeed welcome, as for example if naively plugging the solution (\ref{pureRG1}) within
the complete series at higher orders $\delta^2$ and $\delta^3$, one obtains badly too large values, e.g. $F/\ov\Lam(\delta^2)\sim 0.86$, 
$F/\ov\Lam(\delta^3)\sim 0.96$ for $n_f=2$, principally due to the relatively large coupling $\t \alpha=\pi/2$ in (\ref{pureRG1}). \\

At $\delta^k$-order, Eq.~(\ref{RGred}) is a polynomial of order $k+1$ in $L$, thus
exactly solvable up to third order, with full analytical control of the different solutions.
One can solve both RG and OPT equations numerically, but it also proves particulary convenient to solve both equations
exactly for $L_{RG}(g)$ and $L_{OPT}(g)$, and to look for intersections of those two functions. Note that away from the common intersection solutions, one has
to consider the complete RG Eq.~(\ref{RGstan}) to obtain the right $L_{RG}(g)$ behavior. However, the AF compatibility for $g\to 0$ may be
required either using Eq.~(\ref{RGstan}) or the simpler (\ref{RGred}) with the same results, due to the asymptotic dominance of the relevant terms.  
Whatever the procedure, at increasing $\delta$-orders 
more and more solutions  appear as expected, many being complex in the $\ms$-scheme (complex conjugate solutions since all coefficients of (\ref{OPT}), 
(\ref{RGred}) are real).  Of course not all the solutions are complex at a given order: it is the case at first order $\delta$, but 
at order $\delta^2$ Eqs.~(\ref{RGred}), (\ref{OPT}) actually give 8 different $\t g,\t L$ solutions,   
2 real and 6 complex conjugate ones. Incidentally one of the two real solutions has negative coupling, and the other one a very large $\t \alpha_S\simeq 4$, 
perturbatively completely untrustable. But as motivated above we shall impose the compelling additional constraint that the solutions
should obey AF behavior for $g\to 0$, which already uniquely fixes the critical $a$ value (\ref{acrit}) for the basic interpolation (\ref{subst1}).\\

\begin{figure}[h!]
\epsfig{figure=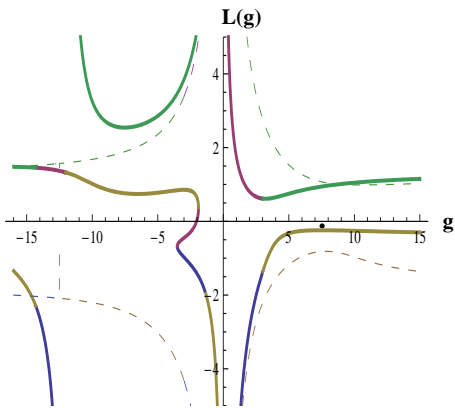,width=6cm}
\caption[long]{The (real parts of) all different branches at order $\delta^3$ (4-loop) of the RG (thick curves) and OPT (dashed curves) solutions 
$L(g)\equiv \ln \frac{m}{\mu}(g)$, in the $\ms$-scheme ($n_f=2$). The (unique) intersection  
solution sitting on both RG and OPT perturbative branches with AF behavior, $L\to -(2b_0 g)^{-1}$ for ${\rm Re}[g]\to 0^+$, occurs at
$g\simeq 7.4 \pm 3.9 i$, $L\simeq -0.23\mp 0.04$, whose real parts 
is indicated approximately by a dot.}
\label{fpirgbranch}
\end{figure}
At least up to the maximal available order $\delta^3$ (four-loop), only a single branch has the right AF perturbative behavior, which provides a 
very clear selection procedure. By simple inspection it is easily checked which solutions are lying on the
unique AF compatible branch.   
For example in Fig.~\ref{fpirgbranch} we plot for the maximal available order $\delta^3$ all the resulting different branches 
of the RG and OPT solutions (\ref{RGstan}), (\ref{OPT}), 
${\rm Re}[\ln \frac{m}{\mu}(g)]$ as functions of the coupling $g=4\pi\alpha_S$, in the $\ms$ scheme. The AF compatible RG and OPT branches clearly appear,  
moreover all other branches are non-ambiguously discardable, having typically the wrong sign for
$g\to 0$, or a grossly different coefficient from the correct AF one $L \sim (-2b_0 g)^{-1}$ for $g\to 0$. 
It thus rejects in this way all spurious solutions of (\ref{RGstan}), (\ref{OPT}) sitting on the other branches, indeed most exhibiting an odd 
behavior, e.g $\t m \ll \ov\Lam$, or $\t g <0$~\footnote{ Note that the pole seen in Fig.~\ref{fpirgbranch} for one of the spurious 
RG branches $L_{RG}(g)$, occuring at $g\simeq -12.02$, is close but slightly different from the perturbative RG fixed point (real) 
solution of $\beta(g)=0$ at four-loop order ($g\simeq -12.96$ for $n_f=2$). 
This shift is due to the fact that $L_{RG}(g)\equiv \ln \frac{m}{\mu}(g) \ne \ln \frac{\ov\Lam}{\mu}(g)$}. 
Incidentally this is the case for the two real solutions 
mentioned above occuring in $\ms$ at order $\delta^2$, or similarly for those seen in Fig.~\ref{fpirgbranch}, actually lying on branches completely disconnected
from the AF compatible ones, thus unambiguously excluded even if one would not know anything about phenomenologically reasonable $F_\pi/\ov\Lam$ values. 
Therefore we pick up the (unique) solution of Eqs~(\ref{RGred})-(\ref{OPT}) sitting on the AF compatible branches at successive orders. \\

 Except for specific approximations explicited below, we perform the optimization mostly using the RG equations (\ref{RGred}) or (\ref{RGgopt}) at 
the naturally consistent perturbative order,
{\it e.g.}  $(k+1)$-loop at order $\delta^k$. Indeed some caution is needed 
with the perturbative order required for RG consistency for the relevant 
$F_\pi$ series, even prior to the OPT modification of the series, due to the subtraction renormalization. 
Perturbative RG invariance typically involves cancellations between contributions from the  beta function and  
the anomalous dimension coefficients $b_k, \ga_k$ of different orders. 
For the  $F_\pi$ series starting with $-s_0/g$, the RG Eq.~(\ref{RGstan}) at order $g^k$ includes consistently 
the $(k+1)$-loop beta coefficient $b_k$: 
$-2b_k g^{2+k} \partial_g (-s_0/g) = (-2 s_0 b_k) g^k$. The RG equation should also incorporate 
more generally the leading and subleading logarithms relevant at a given order $g^k$. \\

One should also be careful to specify the $\ov\Lam$ convention: to compare with most recent other determinations
it will be more convenient to use a standard 4-loop perturbative form~\cite{PDG}, with $b_3\ne 0$:
\bea
&& \ds \ov \Lam^{(n_f)}_4(g) \equiv 
\mu  \,e^{-\frac{1}{2b_0\,g}}\,(b_0\:g)^{-\frac{b_1}{2b^2_0}}  \exp \left[ -\frac{g}{2b_0} \cdot \right. \nn \\ 
&& \left. \left((\frac{b_2}{b_0}-\frac{b^2_1}{b^2_0})   
 +(\frac{b^3_1}{2b^3_0} -\frac{b_1 b_2}{b_0^2} +\frac{b_3}{2b_0}) g\right) \right]\;.
\label{Lam4pert}
\eea
 Now, performing the RG equations (\ref{RGred}) or (\ref{RGgopt}) at order $\delta^k$ rather involves 
the more natural scales at the consistent $k+1$ loop order
$\ov\Lam_{k+1}$, given from (\ref{Lam4pert}) by taking $b_3=0$ (and $b_2=0$) respectively at three-loops (two-loops).
This gives the results shown in Table \ref{tabresms}~\cite{rgoptqcd1}  
\footnote{The results in Table \ref{tabresms} appear slightly different from those in Table 1 of \cite{rgoptqcd1}, as we had
used there rather a (Pad\'e Approximant) 3-loop form for $F/\ov\Lam$, for convenience of comparison with certain lattice
results~\cite{LamlattWilson}.}. Using (\ref{Lam4pert}) or lower orders $\ov\Lam$ for $\t g$ is simply a change 
of normalization convention in the final optimized results, but not changing the optimized values $\t g,\t L$.\\ 

Unfortunately, the unique AF-compatible RG solutions in the $\ms$-scheme in Table \ref{tabresms} remain complex (conjugates). 
As a crude approximation in \cite{rgoptqcd1}, we had considered the differences 
${\rm Re}[F^{(k)}(\t m,\t g)]-F^{(k)}({\rm Re}[\t m], {\rm Re} [\t g])$ 
as indicating a conservative intrinsical theoretical uncertainty. Indeed  
comparing second and first $\delta$-orders in Table~\ref{tabresms}  one observes that
the solution has a much smaller imaginary part. Forgetting momentarily about imaginary parts, 
${\rm Re}\,\t\alpha_S$ decreases to reasonably perturbative values as the $\delta$-order increases.
One may also extract the corresponding $\t m/\ov\Lam$ values at successive orders from Table \ref{tabresms}: this gives
${\rm Re}[\t m/\ov\Lam] \simeq .78, .91, .87$ respectively at orders $\delta$ to $\delta^3$. 
It confirms that the optimized  
mass $\t m$ are well of order ${\cal O}(\ov\Lam)$, although it is partly obscured by the undesirable
imaginary parts. This illustrates our argument discussed  above for expecting a better 
convergence of the RGOPT. \\
At order $\delta^3$, the $g^3 s_4$ term in (\ref{sub}) needs the presently unknown 5-loop
coefficient of $L$. As an approximation, we have estimated $s_4$ either with a Pad\'e Approximant
${\rm PA}[1,2]$ constructed from the lower-order series of coefficients $s_0$ to $s_3$, 
or alternatively simply ignoring this unknown higher-order term, $s_4\equiv 0$, retaining only 
the four-loop RG $\ln^p(m/\mu)$ coefficients. 
The difference between those two choices in Table~\ref{tabresms} 
gives one estimate of higher order uncertainties, and remains very small.\\
\begin{table}[!]
\caption{Combined OPT+RG $\ms$ results at successive $\delta$-order ($n_f=2$ and $n_f=3$).} 
\begin{tabular}{|c||l|l|l|}
\hline
$\delta^k$, RG-order &   $\t L$ & $\t \alpha_S$ & $\frac{F^{(k)}(\t m,\t g)}{\overline \Lam^{n_f}_{k+1}} $ \\
\hline\hline
$\delta$, RG-2l    &       $-0.45 \pm 0.11 i$   &  $1.01 \pm 0.08 i$ & $ 0.266 \pm 0.11 i$\\
\hline
$n_f=3$:           &       $-0.48 \pm 0.09 i$   &  $0.94 \pm 0.18 i$ & $ 0.266 \pm 0.09 i$\\
\hline\hline
$\delta^2$, RG-3l  &      $ -0.52 \mp 0.69 i$     &  $0.73\pm 0.02 i$ &  $ 0.353 \pm 0.03 i$    \\
\hline
$n_f=3$:           &        $ -0.40 \mp 0.19 i$     &  $0.86\pm 0.16 i$ &  $ 0.321 \pm 0.015 i$    \\
\hline\hline
$\delta^3$, $s_4={\rm PA}[1,2]$&   $-0.22 \mp 0.04 i$     & $0.59 \pm 0.31 i$ &   $ 0.349 \pm 0.08 i $       \\
\hline
$n_f=3$:                     &   $-0.22\mp 0.008 i$     & $0.59 \pm 0.30 i$ &     $ 0.356 \pm 0.07i $     \\
\hline
$\delta^3$, $s_4,f_{44}=0$ &     $-0.13\mp 0.04 i$     & $0.61 \pm 0.33 i$ &     $ 0.349 \pm 0.09i $     \\
\hline
$n_f=3$:                      &   $-0.21 \mp 0.008 i$    & $0.60 \pm 0.30 i$ &   $ 0.356 \pm 0.07i $       \\
\hline
\end{tabular}
\label{tabresms}
\end{table}
%
\subsection{Tentative approximate real solutions}
To attempt to cure the problem of non-real solutions, we can try to use again guidance from the GN model. In fact  the  optimized GN mass gap solutions,  
at two-loop order in the $\ms$ scheme, are real for any $N\ge 3$, and a complex solution occurs only for $N=2$~\cite{rgopt1}. 
Then truncating the RG equation (\ref{RGred}) by  
neglecting simply its highest $g^4$ order at $\delta^2$ order (noting that the RG equation for the mass gap at order $g^k$
only needs to hold perturbatively up to terms of order $g^{k+1}$), real solutions were recovered~\cite{rgopt1} for any $N\ge 2$, 
with corresponding results departing from the exact ones by less than a percent. So one expects similarly that truncating the polynomial RG and OPT
equation down to lower orders in the coupling for the $F_\pi$ series will more likely give real solutions. But it does not work so well, as it
requires a cruder RG approximation than for the GN case: 
at the lowest order $\delta$ (two-loop), where the RG Eq.~(\ref{RGred}) is of maximal order $g^3$, real solutions are only recovered when truncating 
this equation down to ${\cal O}(g)$, neglecting all higher orders. The corresponding results are given in Table \ref{tabappcrude} for $n_f=2(3)$.
Similarly at $\delta^2$ (three-loop) order where (\ref{RGred}) is 
of maximal order $g^5$, real solutions are recovered, for $n_f=2$, when truncating the RG equation down to ${\cal O}(g^2)$. 
But the same prescription does not work for $n_f=3$, where the corresponding solution is no longer real. \\
A slightly different approximation can be worked out, by noting that the equivalence
between the RG equation forms (\ref{RGred}) or (\ref{RGgopt}) holds exactly only when the $\ov\Lam$ expression in (\ref{RGgopt}) is used at the same consistent order.
Thus, re-expanding perturbatively Eq.~(\ref{RGgopt}) and truncating it at some lower order will not be fully equivalent to applying directly (\ref{RGred}). 
Real solutions are recovered also in this way, but only at order $\delta^2$ and perturbatively expanding Eq.~(\ref{RGgopt}) up to order $g^2$. 
The corresponding results are also given in the last two lines in Table \ref{tabappcrude} for $n_f=2(3)$.\\
\begin{table}[h]
\caption{Approximated (perturbatively truncated) OPT+RG $\ms$ results at orders $\delta$ and  $\delta^2$, for $n_f=2$ (respectively $n_f=3$). $X$
indicates that no real solutions were found.} 
\begin{tabular}{|l||c|c|c|c|c||c|}
\hline
$\delta^k$, truncation  &   $\t L$ & $\t \alpha_S$ & $\frac{F^{(k)}(\t m,\t g)}{\ov\Lam^{2(3)}_4} $ \\
\hline\hline
$\delta$, $\mbox{RG-2l}_{|{\cal O}(g)}$    &       $-1.31(-1.31) $   &  $0.35(0.36) $ & $ 0.252(0.264) $\\
\hline
$\delta$, [$\partial_g(\frac{F^2}{\ov\Lam^2})]_{|{\cal O}(g^2)}$ &  $-0.99(-1.08) $   &  $0.39(0.387) $ & $ 0.278(0.280) $\\
\hline
$\delta^2$, $\mbox{RG-3l}_{|{\cal O}(g^2)}$  &      $ -0.87(X)$     &  $ 0.47(X)$ &  $ 0.292(X)$    \\
\hline
$\delta^2$, [$\partial_g(\frac{F^2}{\ov\Lam^2})]_{|{\cal O}(g^2)}$    &       $-1.28(-1.43) $   &  $0.38(0.37) $ & $ 0.260(0.257) $\\

                                                                          &       $-0.56(X) $   &  $0.56(X) $ & $ 0.322(X) $\\
\hline
\end{tabular}
\label{tabappcrude}
\end{table}
These different approximate results in Table \ref{tabappcrude} could appear satisfactory at first sight, 
leading to already quite realistic $\ov\Lam$ values, given the crude simplicity of the prescriptions to recover real solutions. 
But there are several problems: first, not surprisingly complex solutions reappear anyway 
at higher order  $\delta^3$, whatever the approximation made, or when varying $n_f$, so that such a prescriction is not robust. 
At order $\delta^2$, if (\ref{RGred}) is truncated at higher orders $g^k, k\ge 3$, 
complex solutions also reappear, close to those in Table \ref{tabresms}. 
Moreover at order $\delta^2$ the second type of approximation gives two real solutions for $n_f=2$, as shown in the last line in Table  \ref{tabappcrude}. 
But an even more serious problem is that
the corresponding solutions $g(L)$ in Table \ref{tabappcrude} no longer satisfy our compelling requirement
(\ref{rgasympt}), they do not have an AF compatible branch. This can be traced to the fact that AF compatibility requires
incorporating all leading logarithms $g^k L^k \sim {\cal O}(1)$, and truncating to too low orders misses some of those,
spoiling the overall RG consistency.
Not only AF compatibility should be considered a crucial underlying requirement, but in practice if no solution
fulfills it there is no really convincing other way to disentangle ambiguous solutions like those appearing for $n_f=2$ at order $\delta^2$. 
When taking either Eq.~(\ref{RGred}) or Eq.~(\ref{RGgopt}) at order $\delta^2$, but now truncating at the next order $g^3$, 
the AF behavior is recovered for a unique solution, but again complex. So these approximations illustrate the generally expected 
incompatibility of AF consistency and (crudely forced) reality. More generally for another perturbative QCD series, or 
in another model, such simple truncations may not even give any real solution at all. 
Even in the simpler GN model, complex solutions occur anyway at higher orders~\cite{rgopt1}. \\
In summary, this calls for an AF compatible, more generic and robust cure to obtain a more reliable determination 
of $\ov\Lam$ or other similar quantities, also fully exploiting the information available from higher order calculations in (\ref{Fpipert}). 
In view of realistic  $\ov\Lam$ and thus $\alpha_S$ determinations, it is also desirable to 
incorporate a convincing manner of estimating intrinsic theoretical uncertainties of the method. 
We will see next that a standard renormalization scheme change allows such a more natural and systematic cure, 
but at the price of a slightly more involved procedure due to the introduction of extra scheme parameters.  
In fact even if the real solutions in Table \ref{tabappcrude} are not strictly AF compatible nor robust, they happen to be numerically 
not far from what will come out below from a more generic and AF compatible prescription. This is because   
they are in smoother continuity with their AF-compatible corresponding solutions at higher orders, in contrast with the grossly inconsistent branches in 
Fig. \ref{fpirgbranch} having typically the wrong sign or a very different coefficient for $g\to 0$, that would give widely different results. 
\subsection{Renormalization scheme changes to real solutions}
\begin{figure}[!]
\epsfig{figure=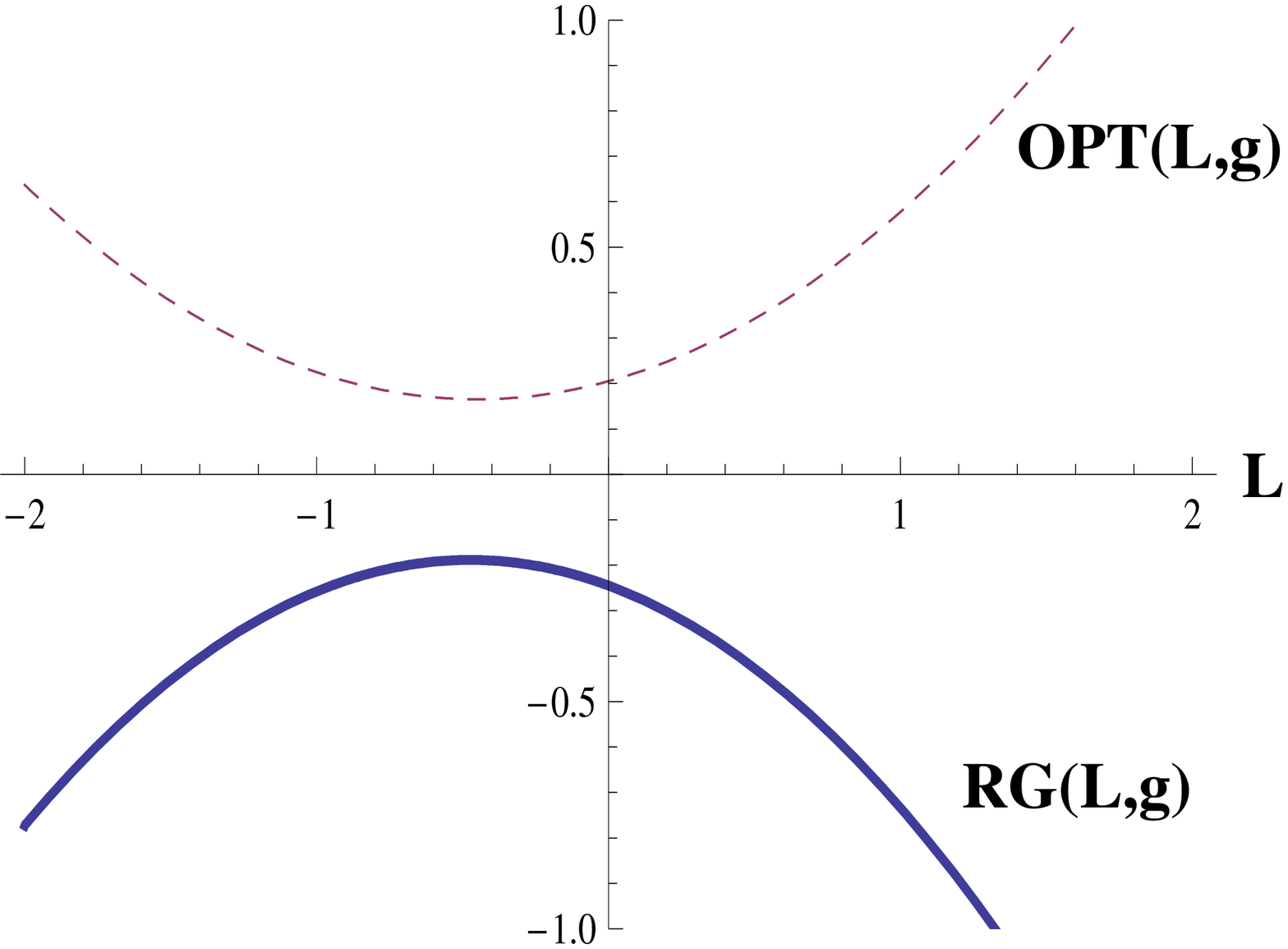,width=5cm}
\vspace{.5cm}
\epsfig{figure=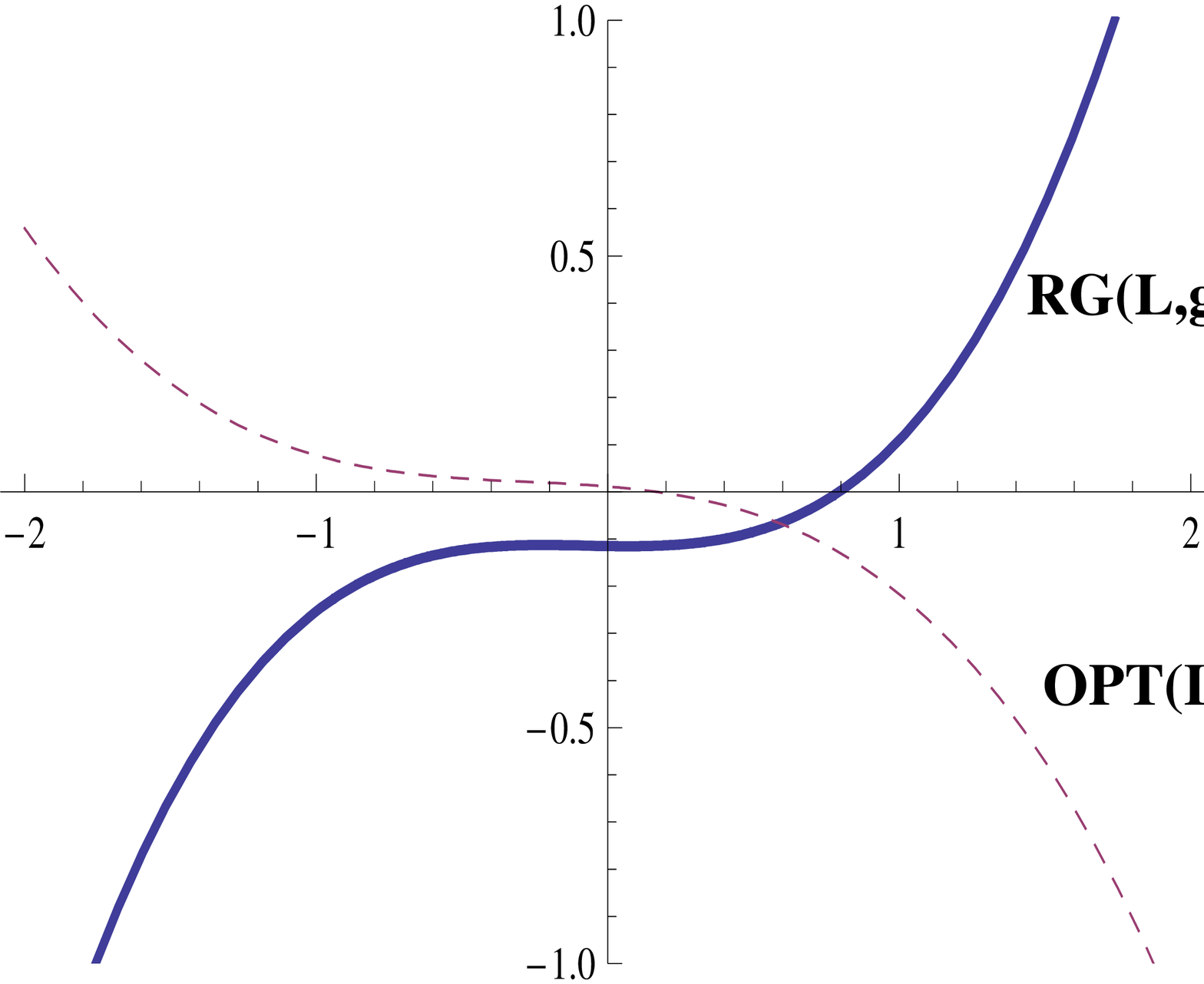,width=5cm}
\caption[long]{RG (thick) and OPT (dashed) Eqs.~(\ref{RGred}), (\ref{OPT}) in $\ms$-scheme ($n_f=2$) as functions of 
$L$ for fixed (real) $g$. Top: order $\delta$, $g\simeq 12.74$; bottom:  order $\delta^2$, $g\simeq 9.15$.}
\label{rgoptbranch}
\end{figure}
Clearly the occurence of non-real solutions, in particular the AF-compatible ones, is simply a consequence of the very familiar fact 
that polynomial equations of order $ \ge 2$ with general real coefficients have complex 
conjugate solutions in general.  For $F_\pi$, and in the $\ms$ scheme, this happens already at 
first $\delta$ order, where the relevant RG and OPT equations are quadratic in $L$, but it will also happen 
in any other scheme or model at increasing orders sooner (most likely) or later.  
Moreover the $\ms$ scheme is particularly convenient and widely used in higher loop QCD calculations, 
and the most standard usage for comparisons of $\Lam$ and $\alpha_S$ results.\\ 
In Fig. \ref{rgoptbranch} we plot the OPT Eq.~(\ref{OPT}) and RG (\ref{RGred})  in the $\ms$-scheme (for $n_f=2$) at orders $\delta$ and $\delta^2$, considered
as functions $L(g)$. Both equations are quadratic (cubic) in $L$ at order $\delta$ ($\delta^2$), and their 
(unique) common root consistent with AF behavior (\ref{rgasympt}) occurs for the complex optimized values $\t g, \t L$ given
in Table \ref{tabresms} (the curves are plotted in Fig.  \ref{rgoptbranch} for real $g\sim {\rm Re}[\t g]$).
 One can see in Fig. \ref{rgoptbranch} that at order $\delta^2$ the OPT and RG equation 
curves cubic in $L$ both have inflection points not far from the real axis (the closest for the OPT curve), respectively at $(-0.28,0.021)$ and 
$(-.075,-.11)$. Thus one can expect that a slight modification of those coefficients  would drive both curves to intersect on the real axis,
so that the corresponding complex conjugate optimization solutions in Table \ref{tabresms} become real. It is also
clear that the imaginary parts of $\ms$ solutions depend much on $n_f$ values. Incidentally, as we will examine in more details 
below, the $\ms$ AF compatible solutions are closer to being real for $n_f=3$, and they even become real (at orders  $\delta$ and $\delta^2$) for $n_f=4$.\\ 

Now to make such perturbations of the various coefficients not an arbitrary deformation, but one fully consistent
with RG properties, a natural and relatively simple prescription is to perform a (perturbative) renormalization scheme change (RSC), defined as 
\be
g  \to g' (1+A_1 g+A_2 g^2+\cdots)
\label{RSCg}
\ee
\be
m\to  m' (1+B_1 g+B_2 g^2+\cdots)
\label{RSCm}
\ee
which can generically affect the coefficients in the defining perturbative series (\ref{Fpipert}) (see also the appendix),   
opening the possibility of real solutions.  However the price to pay is to
introduce more parameters in the procedure, and  various possible such RSC, so we should 
aim at defining a convincing (and possibly fairly unique) prescription. On the other hand the fact that there can be several RSC prescriptions at
a given order can allow firm estimates of theoretical uncertainties of the method, as we will see.  
For an exactly known function of $m$ and $g$, (\ref{RSCg}) or (\ref{RSCm})
would be just changes of variable not affecting any physical result. But for a perturbative series truncated at a given order $g^k$, 
its value in different schemes differs formally by a remnant term of order ${\cal O}(g^{k+1})$.
Accordingly the difference between different schemes is expected to decrease at higher orders, if the coupling 
is sufficiently small. The optimized coupling $\t \alpha_S$ values relevant here for $F_\pi$ are reasonably perturbative 
and decreasing at successive orders, 
but not that small, with $\t \alpha_S \sim {\cal O}(1)$ typically.
Our guideline is thus to recover real solutions if possible, but at the same time with a minimal departure from the original $\ms$-scheme. 
Therefore we will define a {\em minimal} RSC by the following prescriptions:
\begin{itemize}
 \item  i) the RSC incorporates as few extra parameters as possible.
 \item ii) the RSC should also be minimal in the sense of giving a real solution as near as possible to those of the original $\ms$-scheme. 
 \end{itemize}
For i) it appears sensible to consider a change only affecting the mass, Eq.~(\ref{RSCm}).
This is motivated in the OPT framework since the mass is already a trial parameter, and also noticing from Table \ref{tabresms} 
that $\t m$ in the $\ms$ scheme tends to have larger imaginary parts than $\t g$ (at least at orders $\delta$ and $\delta^2$), 
so intuitively it may be more efficient to modify $m$.  
 (Incidentally, we have also tried for completeness to use (\ref{RSCg}), but found no real, AF compatible perturbative solutions at the relevant orders, 
which thus excludes this possibility completely for the $F_\pi$ perturbative series). 
Another practical advantage of (\ref{RSCm}) is that it does not affect the convention chosen for $\ov\Lam$ (see Eq.~(\ref{Lamprime}))~\cite{Lam_rsc}, nor
the RG coefficients $b_i$ in (\ref{RGred}).\\ 

Now, to explore the RSC parameter space 
in such a way that the departure from the original $\ms$-scheme is minimal, for whatever choice in (\ref{RSCm}) 
we will also require ii) above, which can be explicitly controlled by looking for 
the {\em nearest-to-$\ms$ contact} between the RG and OPT solutions considered as two-dimensional curves $\mbox{RG}(g,L)$ and $\mbox{OPT}(g,L)$. \\
 It is also sensible to minimize the amount of extra parameters by considering only one RSC 
parameter at a time in (\ref{RSCm}), essentially (though not necessarily) dictated by the perturbative 
order considered.  At a given $g^k, \delta^k$ order, a naturally preferable RSC parameter is thus one which could give the 
smallest perturbative departure from the $\ms$ scheme, while recovering real solutions. 
For a standard series truncated at order $g^k$ the natural choice is clearly to take $B_k\ne 0$ in (\ref{RSCm}), while a 
lowest order RSC will affect many terms at higher orders, inducing quadratic or higher $B_k$ dependence,  
artificially producing spurious solutions more likely far from the original $\ms$ scheme. Now 
since the relevant $F_\pi$ series (\ref{FpiRG}) starts with $-s_0/g$, $B_k\ne 0$ induces  a quadratic $B_k$ dependence at order $g^k$,  
while a next order $B_{k+1}\ne 0$ gives a linear dependence and is more likely to induce a minimal departure. \\
\begin{figure}[h!]
\epsfig{figure=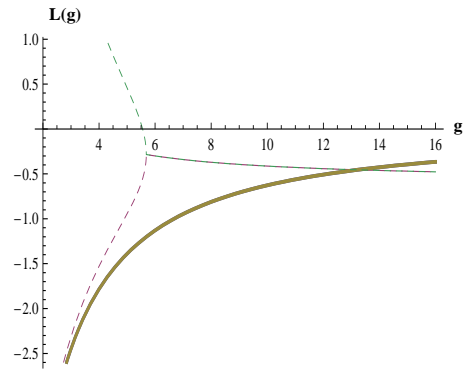,width=6cm}
\caption[long]{The (real parts of the) different RG and OPT solution branches ${\rm Re}[\ln \frac{m}{\mu}(g)]$ at first $\delta$ order (2-loop) in the $\ms$-scheme
(for $n_f=2$). Thick: RG solution; thin dashed: OPT solutions (separating into two real branches below $\t g\lsim  5.7$)}
\label{msbranchd1}
\end{figure}
\begin{figure}[h!]
\epsfig{figure=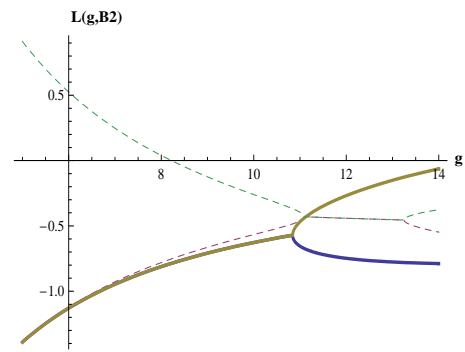,width=6cm}
\caption[long]{Same as Fig.\ref{msbranchd1} at first $\delta$ order for $B_2\simeq 4.48\, 10^{-4}$ corresponding
to the nearest from $\ms$ real contact (for $n_f=2$). Thick: RG solutions; thin dashed: OPT solutions. The contact where both RG and OPT branches are real 
happens at $\t g\simeq 11.06$}
\label{branchd1_contact}
\end{figure}
Such a RSC is illustrated at first $\delta$-order in Fig. \ref{msbranchd1} and \ref{branchd1_contact}: increasing 
$B_2$ from its $\ms$-value $B_2=0$, the OPT and RG curves shown in Fig.(\ref{msbranchd1}) in the $\ms$-scheme, 
move to have a first contact between their real branches for a specific unique $B_2$ value, as illustrated in Fig. \ref{branchd1_contact}. More precisely 
in Fig. \ref{msbranchd1} in the $\ms$-scheme, the (perturbative) branch of the OPT $L(g)$ solution (dashed curves) is real for $g \lsim 5.7$, while
this real branch moves to larger $g$ values for increasing $B_2$ until it reaches the real branch of the RG solution $L(g)$ in Fig. \ref{branchd1_contact} at 
$\t g \sim 11.06$. \\
In practice to fulfill ii) above one does not need to explore the different solution curves in detail, the contact point 
being simply determined analytically by the collinearity of the vectors tangent to the two curves:
\be
\frac{\partial}{\partial g} \mbox{RG} \frac{\partial}{\partial L} \mbox{OPT} -\frac{\partial}{\partial L} \mbox{RG} 
\frac{\partial}{\partial g} \mbox{OPT} \equiv 0
\label{contact}
\ee
where $\mbox{RG}(g,L,B_i), \mbox{OPT}(g,L,B_i)$ designate Eq.~(\ref{RGred}), (\ref{OPT}).
Thus a single RSC parameter is completely determined by the nearest contact requirement (\ref{contact}) 
solved together with (\ref{OPT}) and (\ref{RGred}) determining the corresponding optimized
$\t L$ and $\t g$ values. The combined solutions of the latter three equations for $\t B_i, \t g, \t L$ are easily worked out 
numerically with computer algebra codes~\cite{mathematica}, although a blind numerical solving gives a plethora of spurious
solutions at order $k\ge 2$. Since the RG and OPT equations are both polynomials of order
$k+1$ in $L$, the two equations can alternatively be solved exactly, up to the highest available $\delta^k$, $k \le 3$ (4-loop) order, such that
one can easily find their intersections numerically and crosscheck which solutions appear on the AF-compatible 
perturbative branch solution of behavior (\ref{rgasympt}). The explicit expressions of those exact solutions are however quite involved and not
particularly illuminating, and  anyway do not have a universal physical meaning, resulting from optimization for one particular
RGOPT modified perturbative expansion. Consequently only the value of the RG invariant quantity $F/\ov\Lam$ at their common optimized solution points, 
on the AF compatible branches, should be considered as physically meaningful.  The fact that we are using
the exact $L_{\rm RG}(g)$, $L_{\rm OPT}(g)$ expressions at a given order $\delta^k$ to find real contact/intersection solutions
may be naively interpreted as incorporating some ``nonperturbative'' (in the sole RG meaning) information, keeping in mind that those ``exact'' solutions are actually obtained from 
the (OPT modified) perturbative series, with also perturbative RG acting at a given order. This is justified by trying to use the complete information 
available from the RGOPT modified series at a given order $\delta^k$, since 
the latter is supposed to more efficiently re-sum higher order RG dependence than the original series. We can also   
give the respective perturbative expansions of those solutions for illustration, only used to identify the unique AF-compatible branches.
For instance at order $\delta^2$ in the $\ms$ scheme they read:
\bea
&&L_{\rm RG}(g) =   -\frac{1}{2b_0\,g} +0.93(0.96) +2.8(2.3)\,10^{-2} g \nn \\ 
&&+6.3(1.4)\,10^{-3} g^2 +{\cal O}(g^3);
\label{lrgaf}
\eea
\bea
&& L_{\rm OPT}(g) =  -\frac{1}{2b_0\,g}  +0.44(0.51) -6.5(3.8)\,10^{-3} g \nn \\ 
&& +9.5(-4.5)\,10^{-4} g^2 +{\cal O}(g^3);
\label{loptaf}
\eea
respectively for $n_f=2 (3)$. 
Note that these perturbative expansions should not be confused with the standard perturbative
expansion of $\ln \mu/\ov\Lam$ obtained consistently e.g. up to 4-loop order from (\ref{Lam4pert}), which is different and contains 
also $\ln g$ terms. (The two expressions are trivially related exactly as 
$\ln \mu/\ov\Lam \equiv \ln m/\ov\Lam(g) -L(g)$, and would coincide only 
for $m\equiv\ov \Lam$ exactly, as is the case for the large $N$ limit in the GN model examined in Sec. III).\\

The above considerations favors using $B_{k+1}\ne 0$ in (\ref{RSCm}), but since it is not a compelling prescription, 
we will more conservatively compare the nearest-to-$\ms$ results from a few different RSC choices, 
typically taking $B_2\ne 0$ or $B_3\ne 0$ at $\delta^2$ order, and similarly  $B_3\ne 0$ or $B_4\ne 0$ at $\delta^3$ order.
For $B_{k+1}\ne 0$ at order $\delta^k$ we will generally find a unique nearest-to-$\ms$ real solution, while for $B_k\ne 0$ the quadratic dependence 
often gives two real solutions, but one solution is rejected from being not continuously connected with
the AF branches. We have not explored all possible combinations of RSC prescriptions, but according to the previous arguments our criteria is to reject eventual real solutions
which would give too large $B_k g^k={\cal O}(1)$, not perturbatively trustable and far from $\ms$, while for different RSC choices giving
reasonably perturbative departure from $\ms$, we will include their relative differences within our estimate  
of theoretical uncertainties. \\

 Overall we find that  the nearest-to-$\ms$ contact prescription is very robust for various RSC choices, and also when increasing the $\delta$ order or varying $n_f$. It
gives thus a well-defined and relatively simple procedure to recover real optimized solutions while still 
preserving the consistency of the OPT form (\ref{subst1}) with $a=\gamma_0/(2b_0)$, with compelling RG AF properties of the solutions: 
the critical  $a=\gamma_0/(2b_0)$ is RSC independent, thus not disturbed 
by considering RSC.  For a given RSC the departure from $\ms$ of the optimized solutions can be appreciated by the relative deviations induced in  
the non-universal anomalous dimension coefficients $\ga_i$, $i\ge 1$. Moreover we define 
a ``distance'' from the $\ms$ scheme in the RSC parameter space, simply as
\be
\epsilon_i \equiv \ln (m/m')=\ln (1+B_i g^i)\;.
\label{dmsbar}
\ee
Their corresponding values can be tested for any real solution found, providing quantitative reliability criteria of the optimization results. \\  

Before switching to the concrete results for such RSC, 
we note finally that an alternative prescription could be to optimize with respect to the RSC parameters.  
But it is not guaranteed to give real $\t m, \t g$ solutions, and even if real RSC optimization solutions may be found, they have no reasons to be 
perturbatively close to the original $\ms$ scheme. 
While the mass optimization is the essence of the OPT method, and the additional coupling optimization (\ref{RGred}) or equivalently (\ref{RGgopt}) 
is dictated by RG consistency, a RSC optimization has less compelling motivations. 
We have nevertheless explored this prescription for completeness, but
this RSC optimization is very unstable and not robust, giving either no real solutions or much too large perturbatively unreliable results at increasing $\delta$ orders.
\subsubsection{Nearest-to-$\ms$ RSC: $n_f=2$}
We thus apply the nearest-to-$\ms$ real contact prescriptions at successive orders $\delta^k$, having
performed the substitution (\ref{RSCm}) affecting both the non-logarithmic and RG perturbative coefficients~\footnote{The substitutions (\ref{RSCg}), (\ref{RSCm}) performed in the original
series (\ref{Fpipert}) actually define the {\em reciprocal} RSC $m'\equiv m (1+ \sum_i B_i g^i)^{-1}$, with respect to the one defined in Appendix
in Eq.~(\ref{ga12prime}).} within the original perturbative series  Eq.~(\ref{Fpipert}), prior to modifying it according to (\ref{subst1})
and subsequent RG and OPT optimizations. Note thus that the mass optimization Eq. (\ref{OPT}) is consistently performed in the primed scheme, with 
respect to $m'$, while the (reduced) RG Eq. (\ref{RGred}) is unmodifed for the RSC in (\ref{RSCm}). As already mentioned,  
we will take the full RG dependence available consistently at the order considered, 
which is supposed to incorporate  maximal perturbative information. 
This also guarantees that all the subleading logarithms consistently needed at order $\delta^k$ are taken into account.
For comparison we will also give several results obtained by using different approximations in the RG equation (\ref{RGred})
(as long as those approximations still fulfill the compelling AF compatibility Eq.~(\ref{rgasympt})), to illustrate the
stability of the optimization results.
We show the optimization results at the three successive $\delta^k, k=1,2,3$ orders presently available for $n_f=2$ in Table \ref{tabrealn2} 
 for two different RSC prescriptions with the values of $\t B_k$, $\t B_{k+1}$ giving the nearest-to-$\ms$ real contact. 
\begin{table*}[!]
\caption{OPT+RG solutions with nearest-to-$\ms$ real contact RSC $\t B_i$  at successive $\delta^k$ orders for $n_f=2$. 
The RG Eq.~(\ref{RGred}) is solved including the consistent  $k+1$-loop $b_k$ coefficients in (\ref{rgbi}) at $\delta^k$ order. 
$d_i\equiv (\ga'_i-\overline \ga_i)/\overline \ga_i$  and $\epsilon_i\equiv \ln (1+\t B_i \t g^i) $ (with $i=2\cdots 3$ depending on chosen RSC order $B_i$) 
both measure relative departures from the $\ms$ scheme. Values of other stability/convergence criteria parameters are also given. 
$\ov\Lam_{k+1}$ designates the natural $\ms$ invariant scale, {\it i.e.} defined at the same perturbative $k+1$-loop order for a $\delta^k$-order calculation.} 
\begin{tabular}{|l||l|c|c|c|c|c|c|c|c|c|}
\hline
$\delta^k$, prescriptions &  nearest-to-$\ms$   & $d_i=\frac{\ga'_i}{\overline \ga_i} -1$ &$\ln (1+\t B_i \t g^i) $ &  $\t L'$ & $\t \alpha_S$ & $-2 b_0 \t g \t L'$ &
$\ds\frac{\t m'}{\ov\Lam_{k+1}} $  &$ \ds\frac{F(\t L',\t g)}{\ov\Lam_{k+1}}$ & 
$\ds\mathbf{ \frac{F(\t L',\t g)}{\ov\Lam^{n_f=2}_4 } }$ & $\ov R_{g^{k+1}}$ \\
  &    RSC $ \t B_i $   &                              &                                &         &     &    &        &  &  &    \\
\hline\hline
$\delta$, pure RG-1l   &   $\ms$ ($B_i\equiv 0$)      &  0 & 0 & $-\frac{\ga_0}{2b_0} $   &  $\frac{\pi}{2} $ & 1 & 1  & $\sqrt{\frac{5}{8\pi^2}}$ & $ 0.45$ & \\
\hline
$\delta$,  RG-2l, $B_2\ne 0$ &  $\t B_2=4.48\,10^{-4}$ & $d_2=0.77$ &  $5.3\,10^{-2}$ &   $-0.449 $   &  $0.880$ & 0.61 & 1.22  &0.199 & $\mathbf{ 0.294}$& $+.1$  \\
\hline \hline
$\delta^2$, RG-3l, $B_2\ne 0$ &    $\t B_2=-1.24\,10^{-3} $   &  $d_2=0.6$ & $-8.3\,10^{-2}$ & $ -0.982 $  &  $ 0.638 $  & 0.96 & 0.92  & 0.271 & $ \mathbf{0.269}$& $-.013$   \\
\hline
$\delta^2$, RG-3l, $B_3\ne 0$ & $\t B_3=1.087\,10^{-4} $ & $d_3=-0.78$ & $2.1\,10^{-2}$ &  $ -1.531 $ & $ 0.460 $ & 1.08  & 0.63  & 0.214 & $ \mathbf{0.213}$ & $+.019$  \\
\hline \hline
$\delta^{3*}$, $s_4={\rm PA[1,2]}$ &     $\t B_3=-5.72\,10^{-4} $  & $d_3=4.11$ & $-8.9\,10^{-2}$ &  $-1.483$     & $0.422 $ & 0.96 & 0.72   & 0.2491&  $\mathbf{ 0.2491}$ & $-.007$ \\
\hline
$\delta^3$, $s_4=f_{44}=0$ &  $\t B_3=-5.77\,10^{-4}$  & $d_3=4.16$ & $-9.\,10^{-2}$ &  $-1.476$  & $0.424 $ & 0.96 & 0.72 & 0.2495 &  $\mathbf{ 0.2495}$ & $-.007$  \\
\hline
$\delta^3$, $s_4,f_{44},f_{43}=0$ &  $\t B_3=-4.89\,10^{-4} $  & $d_3=3.53 $ & $-7.6\,10^{-2}$ &  $-1.500$     & $0.422 $ & 0.97 & 0.71 & 0.2460 &  $\mathbf{ 0.2460} $ & $-.006$ \\
\hline
$\delta^3$, $s_4={\rm PA[1,2]}$ &$\t B_4=3.813\,10^{-5}$ & unknown $\overline\ga_4$ &$1.37\,10^{-2}$ & $-1.870$ &$0.347 $ & 0.998 & 0.62 & 0.2224 & $\mathbf{0.2224}$&$+.013$ \\
\hline
$\delta^3$, $s_4,f_{44},f_{43}=0$ &     $\t B_4=3.39\,10^{-5} $  & " " & $1.25\,10^{-2}$ &  $-1.864$   & $0.349 $ & 1.00 & 0.62   & 0.2211 & $ \mathbf{0.2211}$& $+.012$ \\
\hline
\end{tabular}
\label{tabrealn2}
\end{table*}
\begin{table*}[!]
\caption{As in Table \ref{tabrealn2} for $n_f=3$.} 
\begin{tabular}{|l||l|c|c|c|c|c|c|c|c|c|}
\hline
$\delta^k$, prescriptions & nearest-to-$\ms$ & $d_i=\frac{\ga'_i}{\overline \ga_i} -1$ &$\ln (1+\t B_i \t g^i) $ &  $\t L'$ & $\t \alpha_S$ & $-2 b_0 \t g \t L'$ &
$\ds\frac{\t m'}{\ov\Lam_{k+1}} $   &$ \ds\frac{F_0(\t L',\t g)}{\ov\Lam_{k+1}}$ & 
$\ds\mathbf{\frac{F_0(\t L',\t g)}{\ov\Lam^{n_f=3}_4 } }$ & $\ov R_{g^{k+1}}$ \\
  &  RSC $ \t B_i $      &                            &                              &       &       &     &      &  &  &  \\
\hline\hline
$\delta$, pure RG-1l   &   $\ms$ ($B_i\equiv 0$)      &  0 & 0 & $-\frac{\ga_0}{2b_0} $   &  $\frac{\pi}{2} $ & 1 & 1 & $\sqrt{\frac{5}{8\pi^2}}$ & $0.41$ & \\
\hline
$\delta$, RG-2l, $B_2\ne 0$ & $\t B_2=2.38 \, 10^{-4}$ & $d_2=-0.13$ &  $2.1\,10^{-2} $ &   $-0.523 $    & $0.757$  & 0.57 & 1.02  & 0.205 & $\mathbf{ 0.271}$ & $+.07$  \\
\hline \hline
$\delta^2$, RG-3l, $B_2\ne 0$ & $\t B_2=-5.14\,10^{-4} $ & $d_2=0.29$ & $-2.7\,10^{-2} $ &  $ -1.167 $  & $ 0.572 $  & 0.96 & 0.81  & 0.254 & $ \mathbf{0.255}$ & $-.003$  \\
\hline
$\delta^2$,  RG-3l, $B_3\ne 0$& $\t B_3=3.39\,10^{-5}$ &$d_3=-0.32$ & $ 8.7\,10^{-3}$ & $ -1.368 $ & $ 0.507 $  & 0.99  & 0.73  & 0.235  & $ \mathbf{0.236}$  & $+.014$ \\
\hline \hline
$\delta^{3*}$, $s_4={\rm\tiny PA[1,2]}$ &  $\t B_3=-2.52\,10^{-4} $  & $d_3=2.37$ & $-3.7\,10^{-2} $ &  $-1.551$  & $0.416 $ & 0.92 & 0.75  & 0.2546& $\mathbf{0.2546}$& $-.003$ \\
\hline
$\delta^3$, $s_4=f_{44}=0$ &     $\t B_3=-2.53\,10^{-4} $  & $d_3=2.37$ & $-3.7\,10^{-2} $ &  $-1.552$  & $0.416 $ & 0.92 &  0.75  & 0.2545  &  $\mathbf{0.2545}$& $-.003$   \\
\hline
$\delta^3$, $s_4,f_{43},f_{44}=0$ &  $\t B_3=-2.12\,10^{-4} $  & $d_3=1.99$ & $-3.0\,10^{-2} $ &  $-1.573$     & $0.415 $ & 0.94 & 0.73   & 0.2499  &  $\mathbf{0.2499}$&  $-.0025$  \\
\hline
$\delta^3$, $s_4={\rm PA[1,2]}$ &$\t B_4=1.51\,10^{-5}$& unknown $\overline\ga_4$& $ 7.3\,10^{-3}$ & $-1.760$ & $0.374$ &0.94 & 0.70 & 0.2409& $\mathbf{0.2409}$& $+.012$\\
\hline
$\delta^3$, $s_4,f_{43},f_{44}=0$ &  $\t B_4=1.25\,10^{-5}$  & " " & $ 6.3\,10^{-3} $ &  $-1.753$  & $0.377 $ & 0.95 &  0.69   &  0.2389 & $ \mathbf{0.2389}$ & $+.010$ \\
\hline
\end{tabular}
\label{tabrealn3}
\end{table*}
The corresponding relative changes in the anomalous mass dimension $\ga_i$ coefficients at the relevant orders, 
according to Eq.~(\ref{ga12prime}), are also given. 
The values of $\t B_i$ needed to reach a real solution remain overall quite reasonable.  
For example at order $\delta^2$ it gives a relative change on $\ga_2$ or $\ga_3$ of the order of their original $\ms$ values. 
 The amount of the relative distance from $\ms$, $\epsilon_i$ in (\ref{dmsbar}) is also reasonably perturbative, with the best values
obtained for $B_{k+1}\ne 0$ at order $\delta^k$. Notice that the fact that $|\t B_{k+1}|\ll |\t B_k|$
at order $\delta^k$ is largely artificially due to our normalization with $g$ in (\ref{RSCm}): had we used $\alpha_S$ instead to normalize the RSC parameters $B_i$,
these would be $4\pi$ larger and approximately $|\t B_{k+1}|\sim |\t B_k|$. Thus the real quantitative criteria is  the distance from $\ms$, $\epsilon_i$ in (\ref{dmsbar}), 
and the reason for its systematically smaller value for $B_{k+1}\ne 0$ is the systematically lower corresponding optimized coupling $\t \alpha_S$.\\

We give the final ratio $F/\overline \Lam_4$ using  the same 4-loop $\ov\Lam_4$ reference scale expression (\ref{Lam4pert}) 
at the three successive orders, but this is simply a normalization convention convenient for comparison with most other recent $\ov\Lam$ determinations. 
In this normalization $F/\ov\Lam_4$ appears first to substantially decrease from order $\delta$ to $\delta^2$, then slightly reincreasing at 
order $\delta^3$. But the stability is more transparent if using
at order $\delta^k$ the normalization with the scale $\ov \Lam_{k+1}$ 
(taking $b_i =0$ for $i \ge k$ in (\ref{Lam4pert})), which is fully consistent with the RG information actually used 
when performing Eq.~(\ref{RGred}). This is a substantial change at two-loop order $\delta$, and minor at three-loops.  In this more natural normalization the results
at successive orders, indicated in the last before last column in Table \ref{tabrealn2}, in average rather oscillate before stabilizing at orders $\delta^2$ and $\delta^3$,  
which is a good empirical indication of convergence. This behavior is more transparent for $n_f=3$ below. \\
The results for the optimized mass $\t m$,
coupling $\t \alpha_S$, and corresponding optimized $F/\ov\Lam$ values appear empirically convergent. One remarks
the substantial regular decrease of the optimized coupling $\t \alpha_S$ to more perturbative values as the order increases,  correlated
with $\t L <0$ and $|\t L|$ increasing, so that the RGOPT modified sequence becomes more and more reasonably perturbative. (The values of 
$\t \mu/\ov\Lam$, easily obtained from $\t \alpha_S$ using Eq.~(\ref{Lam4pert}), correspondingly increase to more perturbative values, ranging 
from $\t \mu/\ov\Lam \simeq 2.2$, $\simeq 2.36-2.92$, $\simeq 3.15-4$ depending on the RSC prescriptions respectively at orders $\delta$ to $\delta^3$).
We insist however that neither the optimized coupling $\t \alpha_S$, mass $\t m$ or scale $\t \mu$ values have a direct physical meaning, 
being specific results of optimizing the particular series for $F_\pi$. Only the (physical and RG invariant) quantity 
$F/\overline \Lam (\t \alpha_S,\t m)$ will be taken as a prediction.  
In particular the $\t \alpha_S$ values in Table \ref{tabrealn2} (and other Tables below) should not be interpreted as specific $\alpha_S$ predictions, 
but the perturbative behavior of $\t \alpha_S$ and $\t m$ is evidently relevant for the stability/convergence properties of the OPT modified series.
A remarkable feature indeed is the value of $-2b_0 \t g \t L$ given in  Table \ref{tabrealn2}, which at orders $\delta^2$ and $\delta^3$ becomes very close to its 
``maximal convergence'' value $1$ in Eq.~(\ref{rg0gen}) dictated by first RG order dependence. While the optimization including higher-order RG and non-RG terms has 
no reason to fulfill this first order pure RG result, the fact that $\t g \t L$ departs very little from this simplest relation, argued in Sec. III to play an essential
role for the convergence, is another empirical indication of the reliability of the results.  
Finally $\t m/\ov\Lam$ remains relatively close to one in all cases,  as expected from general arguments discussed in Sec. III.
Overall from Table \ref{tabrealn2} the various stability criteria are increasingly satisfied at higher orders $\delta$.
\subsubsection{$n_f=3$}
Since (\ref{Fpipert}) is known for arbitrary $n_f$, we can calculate similarly $F_0/\ov\Lam^{n_f=3}$, where $F_0\equiv 
F_\pi(m_u,m_d,m_s\to 0)$. It is of theoretical interest to compare at this stage the $n_f=2$ and $n_f=3$ results in this strict chiral limit,
where the effects from explicit quark masses are switched off by construction in the RGOPT framework. 
The optimization results for $n_f=3$ are given in Table \ref{tabrealn3}.
We find  a relatively mild variation, as the relevant series coefficients and corresponding optimized solutions are not
strongly dependent on changing from $n_f=2$ to $n_f=3$. But the overall stability and convergence appears  better than
in the $n_f=2$ case. 
 The differences  between different RSC at a given order and corresponding uncertainties, as well as the differences between $\delta^2$ and $\delta^3$
orders, are systematically milder than in the $n_f=2$ case.  This appears to be due 
to the systematically smaller distance to real solutions for $n_f=3$ in the original $\ms$ scheme, as is clear by comparing 
the different values at successive orders $\delta^k$ of the distance to $\ms$ $\epsilon_i=\ln (1+\t B_i \t g^i)$, and deviations induced on $\ga'_i$.  
Basically it originates from the  $n_f$ dependence of the respective RG and OPT polynomial $L$ equations coefficients, 
such that typically at orders $\delta$ and $\delta^2$, the RG and OPT curves in the $\ms$ scheme, like those illustrated for $n_f=2$ in Fig. \ref{rgoptbranch},  
are closer to each other and to the real axis for $n_f=3$, by factors of $2$ to $3$. Thus the required RSC change to reach a real contact solution between 
the two curves is a more perturbative deviation, therefore more stable.  We note that the corresponding real values $F_0/\ov\Lam$ 
(and similarly $F/\ov\Lam(n_f=2)$) are substantially different (systematically lower) 
than what one would obtain by naively taking the real parts~\cite{rgoptqcd1} of the complex results in Table \ref{tabresms}, which is not 
quite surprising since the latter results have relatively large imaginary parts. \\
As a side remark we also mention that the RGOPT results in Tables \ref{tabrealn2}, \ref{tabrealn3} appear more stable and realistic than 
those we had obtained years ago~\cite{qcd1,qcd2} for $F_\pi/\ov\Lam$ from a basically different OPT version, 
briefly recalled above in Sec.~IIB.~\footnote{However, it is opportune to signal   
an unfortunate trivial mistake in ref.~\cite{qcd2} of an overall factor $\sqrt 2$ for $F_\pi/\ov\Lam$, due to a 
confusion with the normalization convention using $f_\pi\equiv \sqrt{2} F_\pi\sim 131$ MeV, affecting Eqs.~(3.6), (3.11), (3.12) of \cite{qcd2}.
The corrected $F_\pi/\ov\Lam$ values in \cite{qcd2} (thus reduced by $1/\sqrt{2}$) appear in retrospect somewhat more 
realistic, although with much larger uncertainties than the present results from RGOPT.}
\subsubsection{Crosschecks from $n_f=4$}
As is clear from previous considerations, the proximity of the
$\ms$  AF-compatible solutions to real ones depends substantially on $n_f$ values. 
Indeed these $\ms$ solutions become real for $n_f=4$ (at orders $\delta$ and $\delta^2$) as we
shortly examine now. Evidently $n_f=4$ is of no direct physical interest since the $SU(4)_L\times SU(4)_R$ chiral symmetry is not
relevant in any crude approximation to the real world, but it indicates useful features and crosschecks concerning the expected accuracy of the RSC prescriptions for
the more relevant $n_f=2, 3$ values.
At first-order $\delta$, the direct optimization results in $\ms$ for $n_f=4$, for the AF compatible branch, has the unique solution:
\be
\t \alpha_S \simeq 0.395\;,\;\; \t L= -1.68\;,\;\;\frac{F_4}{\ov\Lam^{(4)}_4} \simeq 0.200\;,
\ee
and at order $\delta^2$ there are in fact two real solutions sitting on the (unique) AF-compatible branch:
\bea
&&\t \alpha_S \simeq 0.327\;,\;\; \t L= -1.96\;,\;\;\frac{F_4}{\ov\Lam^{(4)}_4} \simeq 0.2276\;, \nn \\
&&\t \alpha_S \simeq 0.657\;,\;\; \t L= -1.54\;,\;\; \frac{F_4}{\ov\Lam^{(4)}_4} \simeq 0.2277\;,
\label{msd2n4}
\eea
with thus very close final results, despite the very different respective $\t \alpha_S$ values. 
The fact that there are two real solutions on the AF branch should not be surprising, since even if $\ms$ solutions are directly real in that
case, there is no reason that for $n_f=4$ these solutions would coincide with an exact contact between the OPT and RG $L(g)$ curves: rather,   
the two solutions correspond to two intersections of those curves. At order $\delta^3$, the AF-compatible solutions become complex in the $\ms$, which 
can be treated with a RSC similar to the ones for $n_f=2, 3$ to recover real solutions. 
Now one may also introduce RSC parameters at order $\delta^2$, 
to explore the domain, perturbatively close to $\ms$, where real AF-compatible solutions occur, 
varying between a contact solution and the $\ms$ intersection solutions (\ref{msd2n4}). These results
vary between $F_4/\ov\Lam^{(4)}\sim .26-.2276$ for $B_2 \sim 2.\, 10^{-4}-0$, and between  $F_4/\ov\Lam^{(4)}\sim .27-.2276$ for $B_3\sim -1.13\, 10^{-5}-0$ 
(where note now that the contact solutions give the real solutions the most distant from $\ms$). The corresponding values of the distances
to $\ms$, $\epsilon_i$ in Eq.~(\ref{dmsbar}), are of the same order as those for $n_f=3$ above. This gives a quantitative crosscheck, knowing in that case
the original $\ms$ real solutions at order $\delta^2$ and controlling at will the distance from $\ms$ with $B_i\ne0$, of what accuracy to expect from the approximated 
real contact solutions. 
\subsubsection{Theoretical uncertainties}
\begin{table}[h!]
\caption{Main optimized values as in Table \ref{tabrealn2} for different approximations on RG dependence ($n_f=2$).} 
\begin{tabular}{|l||l|c|c|c|}
\hline
$\delta^k$, approximation & RSC $ \t B_i $ &  $\t L'$ & $\t \alpha_S$ & $\ds \frac{F(\t L',\t g)}{\ov\Lam^{(2)}_4 } $ \\
 \hline\hline
$\delta$, RG-2l$(g^3=0)$ &   $\t B_2=4.46\,10^{-4}$   & $ -0.40 $     &  $ 1.104 $ &  $ 0.329$  \\
\hline\hline
$\delta^2$,  RG-3l, $s_3= 0$ &    $\t B_2=-1.24\,10^{-3} $  & $ -1.001 $  &  $ 0.631 $  &  $ 0.267$   \\
\hline
$\delta^2$, RG-3l$(g^5=0)$ &    $\t B_2=-1.43\,10^{-3} $   & $ -1.05 $     &  $ 0.618 $ &  $ 0.259$  \\
\hline
$\delta^2$, RG-3l$(g^5=0)$ &    $\t B_3=9.3\,10^{-5} $   & $ -1.452 $     &  $ 0.480 $ &  $ 0.221$  \\
\hline\hline
$\delta^3$, RG-3l($b_3=0$) &    $\t B_3=-4.4\,10^{-4} $   & $ -1.391 $     &  $ 0.443 $ &  $ 0.261$  \\
\hline
$\delta^3$, RG-3l($b_3=0$) &    $\t B_4=3.1\,10^{-5} $   & $ -1.774 $     &  $ 0.363 $ &  $ 0.231$  \\
\hline\hline
\end{tabular}
\label{tabapproxn2}
\end{table}
\begin{table}[h!]
\caption{As in Table \ref{tabapproxn2} for $n_f=3$.} 
\begin{tabular}{|l||l|c|c|c|}
\hline
$\delta^k$, approximation & RSC $ \t B_i $ &  $\t L'$ & $\t \alpha_S$ & $\ds \frac{F_0(\t L',\t g)}{\ov\Lam^{(3)}_4 } $ \\
 \hline\hline
$\delta$, RG-2l$(g^3=0)$ &   $\t B_2=2.4\,10^{-4}$   & $-0.49 $     &  $ 0.87$ &  $ 0.278$  \\
\hline\hline
$\delta^2$, RG-3l, $s_3=0$ &    $\t B_2=-6.3\,10^{-4} $  &  $ -1.09 $  & $ 0.60 $  & $ 0.263$   \\
\hline
$\delta^2$, RG-3l$(g^5=0)$ &    $\t B_2= -4.7\,10^{-4}$   & $-1.14  $     &  $ 0.58 $ &  $ 0.259$  \\
\hline
$\delta^2$, RG-3l$(g^5=0)$ &    $\t B_3= 2.7\,10^{-5}$   & $ -1.29 $     &  $ 0.53 $ &  $0.246 $  \\
\hline\hline
$\delta^3$, RG-3l($b_3=0$) &    $\t B_3= -2.0\,10^{-4}$   & $ -1.46 $     &  $ 0.43 $ &  $0.264 $  \\
\hline
$\delta^3$, RG-3l($b_3=0$) &    $\t B_4\ne 0$: no sol.  & X & X & X\\
\hline\hline
\end{tabular}
\label{tabapproxn3}
\end{table}
Coming back to the real world with $n_f=2,3$, 
we have first studied the stability of the optimization results against some well-defined approximations in the RG dependence (but still preserving the AF-compatibility
of the solutions), or slightly different prescriptions, for both physically relevant $n_f=2, 3$ values.  
More precisely, at  order $\delta^2$ and $\delta^3$  we have  
solved Eq.~(\ref{RGred}) after truncating it by one order in $g$, or neglecting the highest 4-loop RG coefficient $b_3$ at order $\delta^3$.
Also, we studied the effect of neglecting the subtraction term $s_3$ at order $\delta^2$: since $s_k$ appears at order $g^{k-1}$ it should
normally be included for consistency at this order, but strictly speaking $s_k$ is required for the RG equation consistency at the next order $g^k$.  
This is only relevant for the RSC with $\t B_2\ne 0$, since for $\t B_3 \ne 0$ at order $\delta^2$ it only modifies the ${\cal O}(g^2)$ non-logarithmic
terms, so the series only depend on a linear combination $8 B_3+s_3$ (and similarly at order $\delta^3$ when considering only $B_4\ne 0$).
Thus the optimization results taking $B_3\ne 0$ will be identical for $s_3=0$ or $s_3\ne 0$ at order $\delta^2$, with a trivial shift of the unphysical parameter $\t B_3$. 
Indeed neglecting $s_3$ appears to have a minor effect (and similarly if neglecting $s_2$ at the lower order $\delta$).
\footnote{Note that the $s_i$, appearing very small in our chosen normalization 
with $g$ in (\ref{s_i}),  are of the same magnitude as the corresponding non-logarithmic coefficients $f_{ii}$ in (\ref{Fpipert}) at the same orders.}
Overall the optimized RG solutions are very stable with respect to such approximations on the known RG dependence, at the level of a few percent
(see Table \ref{tabapproxn2} and Table \ref{tabapproxn3} respectively for $n_f=2$ and $n_f=3$).
Finally at the maximal available four-loop order $\delta^3$, uncertainties are estimated by taking the unknown $s_4$ coefficient either zero or estimated
by a Pad\'e Approximant PA[1,2]; the difference is almost negligible as one can see in Tables \ref{tabrealn2} and \ref{tabrealn3}. \\

Now a more delicate but important issue is how to estimate realistic theoretical uncertainties of our method, given 
the results in Table \ref{tabrealn2}  and \ref{tabrealn3}, in order to compare with other $\ov\Lam$ determinations 
(and to propagate this theoretical uncertainty in subsequent $\alpha_S$ determinations).
In standard perturbative calculations such uncertainties can be estimated from approximations on the neglected higher orders. 
The very negligible difference in Tables \ref{tabrealn2}, \ref{tabrealn3} at the highest order $\delta^3$ 
between neglecting or approximating the unknown coefficient $s_4$ (that translates into an 
uncertainty of less than 1 MeV on $\ov\Lam$) illustrates an excellent stability at order $\delta^3$, 
but presumably largely underestimates the true theoretical errors of the method. 
On the other hand, the differences at orders $\delta^2$ or $\delta^3$ from truncating the RG dependence in various manners also illustrates stability,
but incorporating these within theoretical uncertainties is unjustified, since in all those cases a part of the available consistent RG information is lost. 
(For example the truncations at order $\delta^2$ with results in Table \ref{tabapproxn2} and Table \ref{tabapproxn3} are actually missing some pieces of
the next-to-next-to-leading logarithmic dependence, normally required at this 3-loop order).\\ 

An advantage of using appropriate RSC to obtain real solutions is that it naturally incorporates reasonably 
convincing uncertainty estimates. First, the differences
between reasonably perturbative but non-equivalent RSC prescriptions at different orders, as illustrated in Tables \ref{tabrealn2} and \ref{tabrealn3}, 
can be expected to give a fair estimate of the theoretical uncertainties. Moreover, 
one can easily quantify their respective departure from the $\ms$ scheme, from the parameters $d_i$ and $\epsilon_i$
given for the different prescriptions in Tables \ref{tabrealn2}, \ref{tabrealn3}.
As motivated previously a linear RSC with $\t B_3\ne 0$ appears more natural at $\delta^2$ order, and indeed fulfills in the best way all stability criteria 
as above explained, concerning the values of $\t \alpha_S$, $\t m/\ov\Lam$, $-8\pi b_0 \t \alpha_S \t L$,  and a minimal perturbative departure from the original
$\ms$-scheme $\epsilon_i$.  
Yet, since the relative departures from $\ms$ for different $\t B_i\ne 0$ are not so drastically  
different, there is no compelling reason to really favor one prescription. We will thus incorporate these RSC differences within a 
conservative theoretical uncertainty estimate, but will also indicate the results obtained if giving 
a stronger preference to the RSC with the minimal departure from $\ms$. All these features are further confirmed from the $n_f=4$ example above, where
the comparison of directly real $\ms$ with the other real solutions obtained from perturbative RSC allows to quantify more precisely the  associated uncertainties. 
 Actually, for the RSC with $B_3\ne 0$ {\em and} $s_4$ approximated with a PA[1,2], we did not find a (real) AF compatible solution at order $\delta^3$, which 
may confirm the intuitively less robust RSC choice with $B_k$ rather than $B_{k+1}$ at order $\delta^k$. We remedied this by 
discarding in this case only the highest order $g^7$ in the RG Eq.~(\ref{RGred}), recovering real (AF compatible) solutions 
given in the sixth lines in Tables \ref{tabrealn2} and \ref{tabrealn3}. Given the very small difference with 
the other results for $s_4=f_{44}=0$ we consider this an acceptable drawback, in order to keep results from different RSC choices to be included within 
uncertainties.\\ 

Complementarily, a crude standard estimate of intrinsic RSC uncertainties may be obtained from the higher order remnant terms, implied
for a given RSC, rooted in the perturbative RSC definition (\ref{RSCm}): when truncating the series at order $g^k$, one
has 
\be 
F_{\ms} (\ov m, g; \ov f_{ij})= F(m', g; f'_{ij}(B_i) ) + g^{k+1} r(B_i)\;.
\label{remnant}
\ee
It is straightforward to estimate such remnant terms
from the original $\ms$ perturbative series expanded to order $g^{k+1}$ for a given RSC in (\ref{RSCm}) and corresponding
real $\t \alpha_S,\t L$ values in Tables  \ref{tabrealn2}, \ref{tabrealn3} (of course 
ignoring the known order $g^3$ in (\ref{Fpipert}), when estimating for instance the RSC remnant at order $\delta^2$). 
The corresponding numbers $ \ov R_{g^{k+1}} \equiv g^{k+1} r(B_i)/\ov\Lam(g) $ are given for illustration in the last column in Tables \ref{tabrealn2}, \ref{tabrealn3},  
giving a rough idea of the expected intrinsic uncertainties if different RSC prescriptions were not available for 
comparison. As expected, these uncertainties decrease substantially as the order increases. Indeed at order $\delta^2$ and $\delta^3$ they lie well within the range spanned by 
the two different RSC prescriptions, bringing these closer (even almost agreeing for $n_f=3$). It is also very consistent with the true 
results at order $\delta^3$, again specially for $n_f=3$. The similar remnant terms at first order $\delta$ are also given for illustration, but are not very useful
since the next order $\delta^2$ is exactly known, indeed they   
overestimate the true uncertainties, better quantified by the smaller differences with the next order $\delta^2$.
(Incidentally, it may seem at first counter-intuitive
that the smaller $|\t B_{k+1}|\ll |\t B_k|$ at orders $\delta^k$ give systematically larger remnant terms $\ov R_{g^{k+1}}$.
As above mentioned the relative $B_{k+1}/B_k$ order of magnitude is artificial, coming from a specific normalization. Moreover   
since $B_k$ is defined one order earlier than $B_{k+1}$, and the RSC cancellations are exact up to order $g^k$, those 
cancellations are still partly operating at the next order $g^{k+1}$ for $B_k$ but not as well for $B_{k+1}$).   
It could be tempting to consider the results corrected by those remnant terms as being closer to $\ms$ (as  
indeed is the case for $n_f=4$ when comparing to the true $\ms$ real solutions). But it is more reasonable to take these as 
only a crude indication of a theoretical uncertainty.
Thus if more conservatively not giving a particular prior to any one of the RSC prescriptions, 
we can take our results as four different  uniformly distributed available predictions at order $\delta^2$ (and six at order $\delta^3$), which allows  
a reasonable averaging procedure. Accordingly we keep also conservatively the full range spanned by the two different RSC (thus with unsymmetrical uncertainties)  
at a given order for the final determination of $F/\ov\Lam$.
For $n_f=2$ such a flat averaging gives:
\be
\ds \frac{F}{\ov\Lam^{(2)}_4}(\delta^2)= 0.243 ^{+.026}_{-.030}\;;\;\;
 \frac{F}{\ov\Lam^{(2)}_4}(\delta^3)= 0.237 ^{+.012}_{-.015}\;.
\label{FLam2fin}
\ee
 Finally we can take the average of the $\delta^2$ and $\delta^3$ results in (\ref{FLam2fin}), 
combining linearly their uncertainties, thus including within the theoretical uncertainties the differences $\ov\Lam(\delta^3) -\ov\Lam(\delta^2)$. 
This is perhaps overconservative, as one may rather naturally take the presumably more precise 
result at the maximal available order $\delta^3$. 
But it is worth keeping in mind that the order $\delta^3$ results partly rely on some approximations  on the unknown coefficient $s_4$, 
 even if this appears very stable. \\

Similarly for $n_f=3$  our theoretical uncertainty is taken as the combination of the differences between different RSC, their intrinsic uncertainties,
and also the orders $\delta^2$ and $\delta^3$ difference. A flat average over the different determinations gives 
\be
\ds \frac{F_0}{\ov\Lam^{(3)}_4}(\delta^2)= 0.248 ^{+.007}_{-.012}\;;\;\;
 \frac{F_0}{\ov\Lam^{(3)}_4}(\delta^3)= 0.251 ^{+.004}_{-.010}\;,
\label{FLam3fin}
\ee
 where like for $n_f=2$, the uncertainties conservatively span the full range of values from the different RSC prescriptions, and the central value is 
averaging over four available values at order $\delta^2$ and six values at order $\delta^3$. The resulting uncertainties are seen to be 
substantially milder than for $n_f=2$, as a direct consequence of the much more stable different results specially at order $\delta^2$. \\
Now a less conservative but logical interpretation of our RSC prescriptions, is to give a stronger ``theoretical prior'' to the RSC prescriptions giving 
the minimal nearest-to-$\ms$ scheme at a given order, being presumably the most perturbative and stable deviation. An extreme such
interpretation simply favors the lower limits in (\ref{FLam2fin})-(\ref{FLam3fin}), which systematically give the minimal distance to $\ms$. But 
perhaps a compromise to put such a prior while keeping a realistic uncertainty estimate, 
is rather to use a weighted average over the different determinations, with (linear) weights  $|\epsilon_i|^{-1}$ simply given by 
the inverse of the well-defined distances to the $\ms$ $\epsilon_i$ for the different RSC prescriptions given 
in Tables \ref{tabrealn2} and  \ref{tabrealn3}. This gives instead of (\ref{FLam2fin}):
\be
\ds \frac{F}{\ov\Lam^{(2)}_4}(\delta^2)= 0.231 ^{+.038}_{-.018}\;;\;\;
 \frac{F}{\ov\Lam^{(2)}_4}(\delta^3)= 0.232 ^{+.017}_{-.010}\;,
\label{FLam2finw}
\ee
and instead of (\ref{FLam3fin}):
\be
\ds \frac{F_0}{\ov\Lam^{(3)}_4}(\delta^2)= 0.246 ^{+.009}_{-.010}\;;\;\;
 \frac{F_0}{\ov\Lam^{(3)}_4}(\delta^3)= 0.249 ^{+.006}_{-.008}\;,
\label{FLam3finw}
\ee
where we conservatively kept the same upper and lower limits, from the unweighted uniform ranges. As one can see these weighted averages tend to give 
slightly lower central values of $F/\ov\Lam$ and $F_0/\ov\Lam$, making the errors less symmetrical for $n_f=2$ but more symmetrical for $n_f=3$. 
For $n_f=2$ it also gives a substantially better agreement between orders $\delta^2$ and $\delta^3$.\\ 

Examining the results in Tables \ref{tabrealn2} and  \ref{tabrealn3} it is clear that the recently
calculated four-loop results for the basic perturbative series~\cite{Fpi_3loop,Fpi_4loop} in (\ref{Fpipert}), that necessitated years of developments,  
is very important to establish empirical stability and convergence
of the RGOPT procedure in this concrete example.  Now, a useful investigation 
is to examine to what extent the final results (especially for  $\ov\Lam^{(3)}$ and $\ov\alpha_S$) 
crucially depend on such involved high-order calculation. To this purpose, we have performed another approximation, applying the RGOPT
at $\delta^3$ order but including solely the three-loop information, {\it i.e.} neglecting all genuine four-loop
coefficients $f_{44}$, $f_{43}$ and $s_4$ in Eq.~(\ref{Fpipert}) (the order $g^3$ coefficients $f_{4i}$ for $i \le 2$ being
entirely determined from lower three-loop results from RG properties). The corresponding results are indicated in Tables  
\ref{tabrealn2},  \ref{tabrealn3}. For both $n_f$ values one can see that neglecting the genuine four-loop contributions has a 
very moderate effect, inducing differences on $\ov\Lam$ of $1-2$\% at most. Though this may be relevant for  
very precise $\alpha_S$ determinations, it further illustrates the very good stability reached at three-loop order, and is a welcome feature 
in view of possible applications in other models with less elaborate higher order calculations.     
It is also worth stressing that already at first non-trivial $\delta$ order, the results in Tables \ref{tabrealn2}, \ref{tabrealn3}  
are quite realistic. Especially for $n_f=3$ it is about less than 
$10\%$ from the average of the orders $\delta^2$ and $\delta^3$ three- and four-loop results.  
The calculation at this $\delta$-order entirely relies on a couple of two-loop graphs as the one illustrated in Fig.~\ref{fpipert},
which is a straightforward calculation.
As argued in section III and supported  by exact results in other models,   
we are confident that this good convergence already at lowest orders is not an accidental feature of the relevant series for $F_\pi$, 
but a more generic property of RGOPT. \\

All results in Tables \ref{tabrealn2},  \ref{tabrealn3} are actually to be understood in the strict chiral limit $m_u, m_d, m_s\to 0$, 
for $F/\ov\Lam^{(2)}$ and  $F_0/\ov\Lam^{(3)}$. 
 It is of interest to compare at this stage the values obtained for $n_f=3$ and $n_f=2$, before switching on the effects from 
explicit chiral symmetry breaking from non-zero quark masses which are evidently more important for $n_f=3$. Then the difference $F-F_0$ only originates
from the $n_f$ dependence in the RG and perturbative coefficients $f_{ij}$, $s_i$ in Eqs.~(\ref{Fpipert}), (\ref{s_i}), 
and clearly $F=F_0$ in the large $N_c$ limit. Thus this difference is expected to be moderate: 
indeed for the pure RG one-loop approximation (\ref{pureRG1}) gives simply 
$F_0/F=e^{-\ga_0/2 (1/b_0(3)- 1/b_0(2))} \simeq 0.97$. At higher orders, 
comparing Tables \ref{tabrealn3} and \ref{tabrealn2} at the same
order $\delta^k$ and for the same RSC, the trend is not very clear, being the result of a rather involved balance between
the $n_f$ dependence itself and the differences it induces in the relevant $\t B_i$, $\t L$, $\t \alpha_S$ values.  
Note that {\em if} the optimized couplings were identical for both $n_f$, 
the ratio $F/\ov\Lam$ would be increasing with $n_f$ (for fixed $F$), since $\ov\Lam$ in (\ref{Lam4pert}) is smaller for $n_f=3$ than for $n_f=2$ for fixed
coupling within the relevant range $\alpha_S\lsim 0.88$, with a larger increase for smaller couplings. This effect is even more pronounced
if $\t \alpha_S(n_f=3) < \t \alpha_S(n_f=2)$ for the same prescription at a given order. But this $\ov\Lam^{(n_f)}$ behavior can be largely compensated by 
the behavior of $F_0/F(\t B_i,\t L,\t \alpha_S)$: for example at order $\delta^2$ and for $B_2\ne 0$, $\ov\Lam^{(2)}/\ov\Lam^{(3)}\sim 1.11$ but 
$F_0/F\sim 0.85$, so that $(F_0/\ov\Lam^{(3)})/(F/\ov\Lam^{(2)}) = .255/.269 \sim 0.95$. Nevertheless the  $n_f=2, 3$ differences 
appear to stabilize at the highest order $\delta^3$, with $F_0/F\lsim 1 $ and $\ov\Lam^{(2)}/\ov\Lam^{(3)}\gsim 1$ but closer to one 
in average over the two different RSC prescriptions. 
Taking the final average values in (\ref{FLam2fin}), (\ref{FLam3fin}), there is a slight tendency that  
$F_0/\ov\Lam^{(3)}$ is smaller than $F/\ov\Lam^{(2)}$ by a few percent, but this is not significant due to the relatively large  $F/\ov\Lam^{(2)}$ RSC uncertainties. 
Similarly, performing as above a flat average over the different RSC results with no particular prior, 
we obtain $F_0/F\simeq 0.99 \pm 0.09$, thus with a relatively large RSC uncertainty mostly due to the one affecting $F$.  
\subsection{Subtracting explicit chiral breaking}
Now, to obtain a more precise determination of $\ov\Lam$, while combining all the relevant results above with theoretical uncertainties,
we must consistently subtract out the {\em explicit}
chiral symmetry breaking effects from 
$m_u, m_d, m_s \ne 0$. In principle it should be possible to incorporate those effects within the variational framework, but 
it would be quite involved, in particular the simplicity of having only polynomial equations in $L$ to solve will be lost, 
so that only numerical solutions could be handled. 
Thus for the time being we will rather rely on other more precisely known results. 
For the $n_f=2$ case, we will take the rather conservative most recent combination of different lattice results~\cite{LattFLAG}:
\be
\frac{F_\pi}{F} \sim 1.073 \pm 0.015.
\label{FpiF}
\ee
although more precise results have been obtained by next-to-leading order ChPT with dispersive analysis: $F_\pi/F =1.0719\pm 0.0052$\cite{FpiFchpt}. 
This non-negligible $\sim 7$\% shift in extracting consistent $\ov\Lam$ values from Eq.~(\ref{FpiF}) is taken into account in the final 
$\ov\Lam^{(n_f=2)}$ numerical values given below, with propagation of the uncertainty in (\ref{FpiF}). 
We obtain
\be
\ov\Lam^{n_f=2}_4 \simeq 359^{+38}_{-26} \pm 5 \mbox{MeV}
\label{Lam2fin}
\ee
where the central value and first errors correspond to the flat averaging discussed above with combined conservative theoretical uncertainties, 
taking values spanned by the different RSC, and the averaging over orders $\delta^2$ and $\delta^3$ results in Table \ref{tabrealn2}. 
The second uncertainty comes directly from the one in Eq.~(\ref{FpiF}).
This additionnal $\sim 1.5$\% uncertainty is smaller, in the $SU(2)$ case, than our conservative theoretical error. In any case
it is legitimate to separate clearly 
the two sources of uncertainties in the final results, as they are unrelated, and improvements in the latter may be foreseen in the future.
Note that if using less conservatively only the results at the maximal available order $\delta^3$ in (\ref{FLam2fin}), one obtains instead
of (\ref{Lam2fin}): $\ov\Lam^{n_f=2}_4 \simeq 363^{+25}_{-17} \pm 5$ MeV.   \\

For $n_f=3$, the effect of explicit chiral symmetry breaking due to the strange quark mass $m_s$  
is clearly much more important than in the $n_f=2$ case, moreover with still relatively large uncertainties, and discrepancies between different 
lattice results~\cite{LattFLAG}. We take the recent available MILC collaboration result in~\cite{FpiF0latt}, which includes ChPT fits 
at the next-to-next-to leading (NNLO) order~\cite{chptNNLO}. The equivalent of the relation (\ref{FpiF}) for $SU(3)$ reads 
\be
\frac{F_\pi}{F_0} \sim 1.172(3)(43)
\label{FpiF0}
\ee
where the first uncertainty is statistical and the last one is systematical.
Our $\ov\Lam$ result for $n_f=3$ is thus indirectly affected by those uncertainties, but this is partly compensated by the overall much more precise
and stable $F_0/\ov\Lam^{(3)}$ in (\ref{FLam3fin}) than the corresponding $n_f=2$ case in (\ref{FLam2fin}). 
From (\ref{FpiF0}) and (\ref{FLam3fin}), we derive
\be
\ov\Lam^{n_f=3}_4 \simeq  317 ^{+14}_{-7} \pm 13 \mbox{MeV}
\label{Lam3fin}
\ee
where the first error is our combined conservative theoretical uncertainty as explained, 
and the second comes from the explicit chiral breaking uncertainty Eq.~(\ref{FpiF0}).
Again if rather using only the results at the maximal available order $\delta^3$ in (\ref{FLam3fin}), one obtains instead
of (\ref{Lam3fin}): $\ov\Lam^{n_f=3}_4 \simeq 315^{+13}_{-5} \pm 13$ MeV.
As one can see, even if  $\ov\Lam^{n_f=2}_4$ is affected by relatively larger uncertainties, there is a clear tendency that $\ov\Lam^{n_f=2} \gsim \ov\Lam^{n_f=3}$.
As discussed above, in the strict chiral limit there is a slight tendency that $F_0/\ov\Lam^{(3)}$ is smaller than $F/\ov\Lam^{(2)}$ by a few percent, but
not quite significantly within theoretical uncertainties, so the effect on $\ov\Lam$ appears mainly due to the explicit breaking in (\ref{FpiF}), 
(\ref{FpiF0}), reducing $F_0$ with respect to $F_\pi$.  \\

At this stage one may compare our results for $\ov\Lam^{n_f=2,3}$ with different classes of lattice calculations based on very different methods, specially for the
$n_f=2$ case not usually accessible from standard perturbative extrapolation.  
First the approach of the Schr\"odinger functional scheme\cite{LamlattSchroed} gave 
$\ov\Lambda^{n_f=2} = 245\pm 16(\mbox{stat})\pm 16(\mbox{syst})$ MeV. But this result has been more recently 
reappraised (mainly due to a substantial change~\cite{Vstatr0new} in the static potential parameter $r_0$ setting the overall physical scales), to  
$\ov\Lambda^{n_f=2} = 310\pm 30$\cite{LamlattSchroednew}. 
Next for Wilson fermions \cite{LamlattWilson}, 
 $\ov\Lambda^{n_f=2} = 261 \pm 17 \pm 26$ MeV. Another approach using   
twisted fermions is based on the determination of the ghost-gluon coupling and also includes 
nonperturbative power corrections in the analysis~\cite{Lamlatttwisted}: the result is
$\ov\Lambda^{n_f=2} = 330 \pm 23 \pm 22 _{-33}$ MeV. Finally another recent independent calculation relies on the evaluation of the static $Q\ov Q$ potential,
with the result~\cite{LamlattVstat}: $\ov\Lambda^{n_f=2} = 315 \pm 30$.
The remaining differences between those different lattice results may be partly related to different dynamical quark mass values in different lattice
calculations, and also different chiral extrapolation methods (see {\it e.g.} the discussion in~\cite{Lamlatttwisted}).
Comparing those results with (\ref{Lam2fin}), our determination appears quite compatible, within theoretical uncertainties, particularly with the
results of \cite{Lamlatttwisted}, although our central value appears somewhat higher.  \\

Concerning $\ov\Lambda^{n_f=3}$, most lattice studies rather present results directly in terms of $\alpha_S(m_Z)$ after perturbative
evolution (see for a recent review \cite{Lamlattrev}), that are thus more appropriately to be compared with our $\alpha_S(m_Z)$ determination 
below, obtained from $\ov\Lambda^{n_f=3}$ in (\ref{Lam3fin}). For instance the HPQCD collaboration obtained\cite{asHPQCD} $\alpha_S(m_Z) = 0.118\pm 0.0008(\mbox{syst})$.  
Apart from other already mentioned lattice determinations entering the latest~\cite{PDG} world average (\ref{asWa}), there is a more recent result for $n_f=3$ from 
the static $Q\ov Q$ potential~\cite{LamlattVstatn3}: $\alpha_S(1.5\mbox{GeV}) = 0.326\pm 0.019$, resulting in $\alpha_S(m_Z) \simeq 0.1156\pm 0.002$.
Finally there are also recent lattice results from the ETM collaboration for $\ov\Lam^{n_f=4}$ including dynamical charm quark~\cite{Lam4latt}, which tend to give
a somewhat higher $\alpha_S(m_Z)$. 
\section{Extrapolating $\alpha_S(\mu)$ at higher scales}
To extract $\alpha_S(\mu)$ at perturbative scales one can now follow a standard procedure: taking as input $\ov\Lam^{(3)}$ for $n_f=3$ (\ref{Lam3fin}),  
one can calculate $\alpha_S(\mu)$ by (perturbatively) inverting Eq.~(\ref{Lam4pert}), at the next (charm quark) threshold scale $\mu\simeq m_c$, 
where one performs the matching using 4-loop (or 3-loop for comparison) perturbative relations~\cite{matching4l} to determine $\alpha_S(\mu\sim m_c)$ and  
corresponding $\ov\Lam^{n_f=4}$. Then we proceed similarly up to $\alpha_S(m_Z)$, having crossed meanwhile the b-quark mass threshold.  From our $\ov\Lam^{n_f=3}$
result we can also provide a determination of $\alpha_S(m_\tau)$, to compare with other independent determinations~\cite{PDG,astau1,alphatau,astau_bj08} 
extracted from the hadronic $\tau$-lepton decays, which have been the subject of intense experimental and theoretical activities 
over the recent years. \\

As far as we are aware, there is no perturbatively safe way of exploiting the $\ov\Lam^{(2)}$ determination to link it to an $\alpha_S$
determination, because of the  intrinsically nonperturbative threshold effects from the strange quark mass $m_s$. In particular, the matching
relations (\ref{match4}) are only valid at scales not too far from heavier quark mass thresholds $\ov m(\mu)$ such that both $\mu\gg\ov\Lam$ and 
$\ov m(\mu)\gg \ov\Lam$ hold. In any cases our $\ov\Lam^{(3)}$ results are more precise than the  $\ov\Lam^{(2)}$ ones, for reasons explained above, which is welcome for
the determination of $\alpha_S$. \\

Explicitly the 4-loop matching relations~\cite{matching4l}, at the same scale $\mu$, read
\bea
&& \ds \alpha_S^{n_f+1}(\mu) = \nn \\
&& \alpha_S^{n_f}(\mu) \left(1-\frac{11}{72} a_s^2
+(-0.972057 +.0846515 n_f) a_s^3  \right. \nn \\
&& \left. +(-5.10032 +1.00993 n_f +.0219784 n^2_f) a_s^4 \right)
\label{match4}
\eea
with $a_s\equiv \alpha_S^{n_f}/\pi$. 
We use for the relevant input the most recent data~\cite{PDG}:
\be
\ov m_c(\ov m_c) = 1.275 \pm 0.025\;;\;\;\; \ov m_b(\ov m_b)= 4.18 \pm 0.03
\label{threshdata}
\ee
where masses are in GeV, and $m_\tau=1.7768$, $m_Z=91.187$. \\
We thus obtain finally:
\be
\ov\alpha_S(m_Z) =0.1174 ^{+.0010}_{-.0005} \pm .0010  \pm .0005_{evol}
\label{asmz}
\ee
and 
\be
\ov\alpha_S^{n_f=3}(m_\tau) =  0.308 ^{+.007}_{-.004} \pm .007 \pm .002_{evol}
\label{astau}
\ee
where again the first uncertainties originate from the RGOPT theoretical uncertainties, as above,
while the second originates from the one in $F_\pi/F_0$ in Eq.~(\ref{FpiF0}). The last evolution uncertainty in (\ref{asmz})  
comes for the remaining matching scale uncertainties, principally around $c$-quark mass threshold, and perturbative RG evolution truncation, 
somewhat similar to typical such estimates when evolving from $\ov\alpha_S^{n_f=3}(m_\tau)$ to $\ov\alpha_S(m_Z)$~\cite{matching4l,PDG,astau1}.\\
We remark that our determination of $\ov\alpha_S(m_Z)$ is rather  
consistent with the most recent world average in (\ref{asWa}), although our central value, relying on (\ref{FpiF0}), is somewhat lower. If considering the 
less conservative interpretation of our RSC prescriptions as discussed above, giving a stronger prior to the really minimal nearest-to-$\ms$ RSC
prescription, the agreement with the world average is improved. One may also compare (\ref{astau}) with the average of various different
determinations, as performed in \cite{PDG}: $\ov\alpha_S^{n_f=3}(m_\tau) =  0.330\pm 0.014$, corresponding to $\ov\alpha_S(m_Z) =0.1197\pm 0.0016$, although 
some recent studies with an improved treatment of nonperturbative effects find slightly lower values~\cite{astau_lower} 
(see also the discussions and various approaches in \cite{alphatau,astau_bj08,astau_beneke_recent}). \\
However, one should keep in mind that our central $\ov\alpha_S$ value sensitively 
depends on the $F_\pi/F_0$ value (\ref{FpiF0}), relying on the analysis in \cite{FpiF0latt} with NNLO ChPT~\cite{chptNNLO} fits, and somewhat higher or lower
values have been obtained by other lattice collaborations~\cite{LattFLAG}. 
As discussed {\em e.g.} in \cite{LattFLAG}, the overall present status of lattice results for parameters related to the chiral $SU(3)$ limit  
is not quite clear due to difficulties in extrapolating to the exact chiral limit, with a hinted  
slower convergence of $SU(3)$ ChPT. Indeed there can be important cancellations between the NLO and NNLO ChPT contributions~\cite{FpiF0chpt,chptNNLO} to $F_\pi/F_0$.   
Also it has been theoretically advocated~\cite{rechpt} and more recently reanalyzed in \cite{rechpt_recent} 
that $F_0$ could be substantially more suppressed than $F$ due to the special role of the strange quark mass in the $SU(3)$ ChPT case. 
Thus if $F_\pi/F_0$ happens to be somewhat larger (smaller) with better accuracy
in the future, clearly our $\ov\alpha_S$ central value in (\ref{asmz}) will be correspondingly smaller (larger) (approximately by $\sim \mp .0004$ for a shift
of $\pm .02$ in $F_\pi/F_0$). \\
Now since our calculation is basically a determination of $F_0/\ov\Lam$, one may use it alternatively, taking the world average 
$\alpha_S$ value in (\ref{asWa}) as input for the RGOPT predictions (\ref{FLam3fin}) to determine $F_\pi/F_0$.  This gives:
\be
\frac{F_\pi}{F_0} \simeq 1.12 ^{+.05}_{-.025} |_{\mbox{th,rgopt}} \pm .03_ {\alpha^{w.a.}_S} \pm .02_{evol},
\ee
with the uncertainties from different origin separately shown (the first upper bound corresponding to the really minimal nearest-to-$\ms$ RSC prescription).   
%
\section{Conclusions and prospects}
In this paper we have developed a rather straightforward implementation of RG properties
within a variant of the variationally optimized perturbation, initiated in \cite{rgopt1,rgoptqcd1}, using only perturbative information 
at the first few orders. The role of RG in the OPT for renormalizable models was hinted to long ago~\cite{gn2,qcd1,qcd2} 
but restricted to the linear interpolation $a=1$ case in (\ref{subst1}), and in a previously not very transparent formalism difficult to handle and to generalize. \\

To summarize the main results and properties in our present approach:\\
i) first, a crucial difference with most previous OPT considerations, is that the compelling AF compatibility requirement (\ref{rgasympt})
with the perturbatively modified RG properties generically and uniquely fix the variational mass interpolation (\ref{subst1}) in terms 
of the {\em universal, first order} RG coefficients: $a\equiv \ga_0/(2b_0)$. This result is valid at arbitrary orders, since higher RG orders 
and non-logarithmic coefficients are all subleading as far as the AF behavior is concerned.\\
ii) Furthermore this is demonstrated to improve drastically the convergence and stability of the method, as supported by exact results in the
GN model, and more generically by exact results in the pure first RG order approximation for an arbitrary AF model, as shown in Sec. III.C, with maximal
convergence in this case.\\
iii) For a general perturbative series at higher orders, due to the dependence on higher RG order and non-RG coefficients 
the optimized RG and OPT solutions depart from the generic first RG order ones (\ref{rg0gen})-(\ref{opt0gen}), 
but due to the previous properties both the RG and OPT equations have a unique AF-compatible branch solution. 
Up to the highest available (4-loop) order for $F_\pi$, the solutions and their properties are fully analytically controllable, from 
polynomial equations up to fourth order, and their common solutions do not depart much from the first RG order relation (\ref{rg0gen}), 
with the welcome property that the optimized coupling $\t \alpha_S$ appears empirically to rapidly stabilize towards more ``perturbative'' values. 
However those solutions are non-real in general, resulting from solving exactly polynomial equations of order $ \ge 2$ in $L\equiv\ln m/\mu$ with general real coefficients.  
For the relevant perturbative series for $F_\pi$, and in the $\ms$ scheme, this happens already at first order $\delta$ (where the relevant equations 
are quadratic in $L$, but it will also happen for other quantities in any other scheme or model at increasing orders.\\
iv) To recover real physical optimized solutions, we thus proposed a simple cure,
by considering perturbative renormalization scheme changes (RSC), at the price of introducing more parameters, however fixed
by the well-defined criteria (\ref{contact}) of the nearest-to-$\ms$ RSC. For the obtained solutions the discrepancies from the original $\ms$ scheme are 
seen to decrease consistently at increasing orders, as expected from general RSC arguments, and provide also a well-defined measure of 
the theoretical uncertainties of the method.\\
v) In practice our $F_\pi/\ov\Lam$ calculation at first $\delta$-order gives already very reasonable approximations, 
 about 10\% from realistic determinations from other nonperturbative methods, principally lattice ones.
At second and third orders our results exhibit a remarkable stability, and the RGOPT accuracy obtained, with conservative uncertainty estimates, 
gives a new independent theoretical $\alpha_S$ determination  well comparable to the most precise recent {\em single} determinations of $\alpha_S$, 
including some very recent lattice determinations with fully dynamical quarks. We expect this not to be a mere accidental feature of the relevant $F_\pi$
series, but a more generic property rooted in the remarkable RG resummation shortcuts operated by the RGOPT.\\

The mostly used linear version of the OPT, with $a=1$ in (\ref{subst1}), often gives at the first orders empirically reasonable nonperturbative approximations 
beyond the large $N$ or mean field approximations in renormalizable models, but due to technical limitations most of those studies  
were restricted until now to the first or second OPT orders in realistic $D > 2$ models, specially when including finite temperature effects
~\footnote{An exception is the so-called ``screened perturbation theory'' or ``Hard Thermal Loop resummation''~\cite{HTL}, closely related to the OPT basic idea~\cite{delta}
in the thermal QCD context, where calculations up to three-loops have been
performed~\cite{HTL3}, although mainly in the high temperature expansion approximation.}. We have seen that for the GN model the linear OPT is recovered consistently strictly
in the large $N$ limit, and we suspect this may be the case for a large class of models with a $O(N)$ or related symmetry, 
although we have not checked this systematically (it amounts simply to check the values of $ \ga_0/(2b_0)$ or its equivalent for large $N$, whenever it is defined).
(Note however that in QCD  $\ga_0/(2b_0)\ne 1$ for large $N_c$: $\ga_0/(2b_0)\to 9 (N_c^2-1)/(22 N^2_c)\to 9/22 $ for $N_c\to \infty$, but 
the large $N_c$ limit in QCD involving planar graphs is evidently more involved 
than for simpler $O(N)$ models; an investigation in the OPT context is beyond our present scope). 
Beyond the large-$N$ approximation, and for models having both non-trivial mass and coupling renormalizations, 
our results strongly indicate that the linear OPT is expected to converge very slowly at higher orders, or even not to converge,  
for certain quantities related to composite operators with extra renormalization, like the series used here for $F_\pi$ typically.\\

A  more general and deeper unsolved question is whether the OPT or RGOPT version really ultimately 
converges at higher orders in a renormalizable theory. 
It appears unlikely that the relatively simple trick of reshuffling the interaction terms with a variational mass could really 
circumvent the well-established fact that in higher dimensional renormalizable models
the perturbative series are very likely only asymptotic, due to the factorially divergent perturbative coefficients at high orders 
with an associated essential singularity at the origin in the complex coupling plane~\cite{gzj_lopt,renormalons}. While the
convergence of our prescription is generic and maximally fast for the pure RG dependence (as explicitly shown 
in Sec. III.C), it is also known that the higher order perturbative singularities in an AF theory
like QCD, the infrared renormalons, are actually originating from the non-RG dependence of the perturbative series at high orders~\cite{renormalons}. The (linear) 
$\delta$-expansion for $a=1$ was shown~\cite{Bconv} to damp those renormalons at intermediately large orders, 
but the factorial growth of coefficients appear to ultimately overcompensate this behavior. 
However, in principle asymptotic series have a minimal truncation error at some order, difficult to determine in general except for very simple models.  
What we can conjecture here, is that even if probably not ultimately convergent, the RGOPT AF-compatible version may be such that the optimal truncation
order is relatively low, as is empirically supported by our results for the relevant $F_\pi$ series, where the 
stability clearly happens at order $\delta^2$ to $\delta^3$. A more general investigation of this conjecture is evidently beyond the scope of the present work and left
for the future. \\

Leaving aside those general considerations, one can have more pragmatic prospects, first by    
investigating further the physics of dynamical chiral symmetry breaking in QCD. Our next logical step will be to estimate from our RGOPT version  
the other main chiral symmetry breaking order parameter, the quark condensate. We also plan to estimate some of 
the low-energy constants of ChPT~\cite{chpt}, at least in a first stage 
the few ones which are directly accessible in the exact chiral limit. To go further our approach needs to 
implement consistently the explicit chiral symmetry breaking terms from the $u, d, s$ quark masses (which would allow in particular 
to address the evaluation of $F_\pi/F$, $F_\pi/F_0$ directly  in terms of the light quark masses within the RGOPT, to be compared with ChPT predictions~\cite{chpt,chptNNLO}).   
Incorporating those effects within the OPT framework is formally straightforward,  
but the whole RGOPT procedure will become more involved, typically the simple form of the RG and OPT equations (\ref{RGred}), (\ref{OPT}) polynomial in $L$ will be lost, 
so that only purely numerical solutions may be available. 
%
\appendix
\section{RG and Perturbative expressions}
\subsection{RG conventions and expressions}
Our normalization convention for the beta-function and anomalous mass dimension entering the
standard RG operator in Eq.(\ref{RG}) is 
\be
\beta(g)\equiv \frac{dg}{d\ln\mu} = -2 g^2 \sum_{i\ge 0} b_i g^i
\label{beta_conv}
\ee
\be
\gamma_m(g)\equiv -\frac{d\ln m}{d\ln\mu} = g \sum_{i\ge 0} \gamma_i g^i\;,
\label{ga_conv}
\ee
where  for QCD $g\equiv 4\pi \alpha_S$.
The QCD $b_i$ and $\gamma_i$ coefficients, known up 
to four loops~\cite{bgam4loop}, read in the $\ms$ scheme in our conventions
\bea
&& (16 \pi^2)\,  b_0 = 11-\frac{2}{3}n_f \nn \\
&& (16 \pi^2)^2\, b_1 = 102-\frac{38}{3} n_f \nn \\
&& (16 \pi^2)^3\,  b_2 = \frac{2857}{2} -\frac{5033}{18} n_f +\frac{325}{54} n^2_f \nn \\
&& (16 \pi^2)^4\,  b_3 = (\frac{149753}{6}+3564 z_3)-(\frac{1078361}{162} +\frac{6508}{27} z_3) n_f \nn \\
& & +     (\frac{50065}{162}+\frac{6472}{81} z_3) n^2_f +\frac{1093}{729} n^3_f 
\label{rgbi}
\eea
and
\bea
&&  \gamma_0  = \frac{1}{2\pi^2} \nn \\
&& (16 \pi^2)^2\, \gamma_1 = \frac{404}{3}-\frac{40}{9} n_f \nn \\
&& (16 \pi^2)^3\, \gamma_2 = 2 (1249 -(\frac{160}{3} z_3 +\frac{2216}{27}) n_f -\frac{140}{81} n^2_f) \nn \\
&& (16 \pi^2)^4\, \gamma_3 = 2 \left(\frac{4603055}{162} +\frac{135680}{27} z_3 -8800 z_5 \right. \nn \\ 
&& \left. +(-\frac{91723}{27}-\frac{34192}{9} z_3 +880 z_4 +\frac{18400}{9} z_5) n_f \right.\nn \\ 
&& \left.
 +(\frac{5242}{243}+\frac{800}{9} z_3-\frac{160}{3} z_4) n^2_f+ (-\frac{332}{243}+\frac{64}{27} z_3) n^3_f \right) \nn \\
\label{rggai}
 \eea
where $z_i\equiv\zeta(i)$ are the Riemann Zeta functions.\\

For a general perturbative change of renormalization scheme, according to
\be
g \to g' \equiv f_g(g) = g (1+A_1 g+A_2 g^2+\cdots)
\label{RSch}
\ee
\be
m\to m' \equiv m\: f_m(g) = m (1+B_1 g+B_2 g^2+\cdots)
\ee
the expressions of the RG coefficients in the primed scheme in terms of the original scheme are easily obtained at arbitrary orders 
from RG properties, resulting formally in 
\be
\beta'(g') = \beta(g) (1+g \frac{\partial}{\partial g})f_g(g)\;,
\label{betprimegen}
\ee
\be
\ga'_m(g') = \ga_m(g) -\beta(g) \frac{\partial}{\partial g} \ln f_m(g)\;.
\label{gamprimegen}
\ee
As is well-known $b_0$, $b_1$ and $\ga_0$ are scheme-invariant, and for the first few higher-order coefficients, explicitly one has:
\bea
&& b'_2=b_2-A_1 b_1+(A_2-A^2_1)b_0\,, \nn \\
&& b'_3=b_3-2A_1 b_2+A^2_1 b_1 +(4A^3_1-6 A_1 A_2+2A_3)b_0 \nn \\
\label{b2prime}
\eea
\bea
&& \ga'_1 = \ga_1+2b_0 B_1 -\ga_0 A_1\,,\nn \\
&& \ga'_2 = \ga_2+2B_1 b_1 +2 (2B_2-B^2_1)b_0 -\ga_0 A_2-2A_1 \ga'_1\,,\nn \\
&& \ga'_3 = \ga_3 -A_3 \ga_0 -(A^2_1+2A_2)\ga'_1-3A_1 \ga'_2 \nn \\ 
&& +2b_0 (B^3_1 -3B_1 B_2 +3B_3)-2b_1(B^2_1-2B_2)+2b_2 B_1 \nn \\
\label{ga12prime}
\eea
and so on. A well-known consequence~\cite{Lam_rsc} is that for an AF theory, $\ov\Lam$ to any order is only affected
by the first order term in (\ref{RSch}):
\be
\Lam' \equiv  \ov\Lam \:\exp{(\frac{A_1}{2b_0})}
\label{Lamprime}
\ee
\subsection{Perturbative $F_\pi$ and related expressions}
The original standard perturbative series for $F_\pi$ up to $\alpha_S^3$ four-loop order Eq.~(\ref{Fpipert}) may be written as 
\bea 
& F^2_0 (pert) = 3 \frac{m^2}{2\pi^2} \left[ \mbox{div}(\epsilon,\alpha_S) +f_{10} L +f_{11}  \right. \nonumber \\ 
& \left. +\frac{\alpha_S}{4\pi}(f_{20} L^2 +f_{21} L +f_{22}) \right. \nonumber \\ 
&\left. +(\frac{\alpha_S}{4\pi})^2 (f_{30} L^3+f_{31} L^2 +f_{32}L +f_{33}) \right.\nonumber \\
&\left. +(\frac{\alpha_S}{4\pi})^3 (f_{40} L^4+f_{41} L^3 +f_{42}L^2+f_{43}L +f_{44})
\right]  \nn \\ \label{Fpipapp}
\eea    
where $L\equiv \ln \frac{m}{\mu}$ and $m$ is the $\ms$ scheme running mass.   
The coefficients of the $L$ terms and of the non-logarithmic terms
at successive $\alpha_S$ order, with explicit $n_f$ dependence entering at $\alpha^2_S$ three-loop order, are given in this normalization as 
\bea
&& f_{10}= -1,\;\;  f_{11} = 0\;, \nonumber \\
&& f_{20} = 8,\;\;  f_{21} = \frac{4}{3}, \;\;f_{22}= \frac{1}{6}\,,\nonumber \\
\eea
\bea
&& f_{30}= -\frac{304}{3} +\frac{32}{9}n_f,\;\;  
f_{31} = \frac{136}{3} -\frac{32}{9}n_f, \nonumber \\
&& f_{32} = 8 z_3 +\frac{149}{9} +\frac{10}{9} n_f, \nonumber \\
&& f_{33} = -\frac{53}{3}-\frac{4}{3}B_4 +6 z_4 +\frac{80}{3} z_3 -n_f (\frac{14}{9}z_3-\frac{25}{9})\nn \\
\eea
\bea
&& f_{40}= -\frac{16}{27}(57 - 2n_f)(45 - 2n_f), \nonumber \\ 
&& f_{41} = \frac{32}{81}(3438 -441 n_f + 8n^2_f), \nonumber \\
&& f_{42} = \frac{4}{81} (-4977 + 4860z_3 + 2n_f (897 + 20 n_f + 972z_3)), \nonumber \\
&& f_{43} =  \frac{1}{1458} \left( (20736a_4 + 864\ln^2 2(\ln^2 2-\pi^2)) (-45 + 2n_f) \right. \nn \\
&&  \left. + 4n_f (47181 - 2345n_f - 198\pi^4 + 
     54(-507 + 38n_f)z_3) \right. \nn \\
     && \left. + 9(-98839 + 900\pi^4 + 95304z_3 + 28080z_5) \right),  \nonumber \\
&& f_{44} = -\frac{25.62696915 -8.147150256 n_f -0.08246295965 n^2_f}{4} \nn \\
\eea
with $B_4= 16 a_4 +\frac{2}{3} \ln^2 2(\ln^2 2 -\pi^2) -\frac{13}{180} \pi^4 \sim -1.762800087$, $a_4= {\rm Li}_4(1/2)$.   
Note that apart from the four-loop non-logarithmic terms $f_{44}$ known only in numerical (but accurate) form, all other coefficients 
have been calculated fully analytically for arbitrary colors $N_c$, $n_h$ massive and $n_l$ massless quarks~\cite{Fpi_3loop,Fpi_4loop}. (Here
we took consistently $n_h\equiv n_f$, $n_l=0$ since the mass dependence of the $n_f$-degenerate lightest quarks
plays a determinant role in the optimization.) \\
For completeness we also give explicitly the divergent contribution $\mbox{div}(\epsilon,\alpha_S)$ 
remaining after mass and coupling renormalization in the $\ms$ scheme~\cite{Fpi_3loop,PrivComm}. 
Using dimensional regularization with  $D\equiv 4-\epsilon$ in our normalization, it reads
\bea
&& \mbox{div}(\epsilon, \alpha_S) = \frac{1}{\epsilon} +\frac{\alpha_S}{4\pi} (-\frac{8}{\epsilon^2}+\frac{10}{3\eps}) \nn \\
&& +(\frac{\alpha_S}{4\pi})^2 (\frac{-24z_3 +\frac{910}{3}-16n_f}{9\eps}+ \frac{-132 +\frac{40}{9}n_f}{\epsilon^2} -\frac{f_{30}}{\epsilon^3} ) \nn \\
&&  +(\frac{\alpha_S}{4\pi})^3 (\frac{X_3}{\eps}+\cdots)
\label{Fpidiv}
\eea
where the most relevant ${\cal O}(\alpha^3_S)$ simple pole $\eps^{-1}$ term is
\bea
&& X_3 = \frac{144737}{324} +\frac{\pi^4}{3} -\frac{9770}{27} z_3 +\frac{260}{3} z_5 \nn \\ 
&& +\frac{n_f \left(92595 -29210 n_f -1188 \pi^4 +1620 (79 + 2 n_f) z_3 \right)}{7290} \nn \\
\eea
(the higher $\eps^{-p}$ powers being as usual determined from lowest orders from RG properties).\\
To determine the subtraction terms in Eq~(\ref{sub}), the simplest way, as explained in the main text,
is to find the remnant of the action of the homogeneous RG operator (\ref{RG}) on the finite part of (\ref{Fpipapp}), 
and to match it order by order with the RG acting on (\ref{sub}) to determine the coefficients $s_i$.
Using RG properties (on those finite expressions) as usual at a given order $g^k$ the $f_{k,k-1}$ coefficient of the $\ln m$ term and the non-logarithmic $f_{kk}$ coefficient
have to be calculated, while other leading and subleading logarithmic coefficients $f_{k,0}, \cdots, f_{k,k-2}$ are entirely determined by
lower orders. More precisely, rewriting the finite part of Eq.~(\ref{Fpipapp}) in the more convenient normalization:
\be
F^2_\pi \equiv m^2 \sum_{k=0}^3 g^k \sum_{p=0}^{k+1} a_{k+1,p}[\ln (\frac{m}{\mu})]^{k+1-p}
\ee
with straightforward connection between the $a_{kp}$ and $f_{kp}$ coefficients of (\ref{Fpipapp}) ($a_{10}=-3/(2\pi^2)=3 f_{10}/(2\pi^2)$, etc),  
one obtains the following useful relations:
\bea
&& a_{20}= -\ga_0 a_{10}; \nn \\
&& a_{21} =  (\frac{\ga_1-b_1}{\ga_0-b_0})a_{10}  -\ga_0 (a_{10}+2a_{11}).
\eea
\bea
&& a_{30}= -\frac{2}{3}(\ga_0+b_0) a_{20};\nn \\
&& a_{31}= -\ga_1 a_{10} -\ga_0 a_{20} -(b_0+ \ga_0) a_{21}; \nn \\
&& a_{32} = (\frac{\ga_2-b_2}{\ga_0-b_0})a_{10} -\ga_1 (a_{10} +2 a_{11}) -\ga_0 a_{21}\nn \\
&& -2(b_0+ \ga_0) a_{22}.
\eea
\bea
&& a_{40}= -(b_0+\ga_0/2) a_{30}; \nn \\
&& a_{41}=  -\frac{2}{3}(b_1+ \ga_1) a_{20} -\ga_0 a_{30}
 -\frac{2}{3}(2b_0+ \ga_0) a_{31};\nn \\
&& a_{42}= -\ga_2 a_{10}- \ga_1 a_{20} -(b_1+ \ga_1) a_{21} -\ga_0 a_{31}\nn \\
&& -(2b_0+ \ga_0) a_{32}; \nn \\
&& a_{43}= (\frac{\ga_3-b_3}{\ga_0-b_0})a_{10} -\ga_2 a_{10} -2 \ga_2 a_{11} -\ga_1 a_{21}\nn \\
&& -2(b_1 + \ga_1) a_{22} - \ga_0 a_{32} -2(2b_0 + \ga_0) a_{33}.
\eea
Including the subtraction terms $s_i, i\ge 1$ as defined in Eq.~(\ref{sub}) 
simply changes the non-logarithmic coefficients in those relations by $a_{ii}\to a_{ii}-s_i$. 
The $s_i$ can be expressed in terms of RG coefficients, but beyond the first two terms it is not particularly transparent. They read explicitly:
\bea
&& s_0 = \frac{3}{4\pi^2(b_0-\gamma_0)},\;\; 
s_1= 3\left(-\frac{5}{24\pi^2} +\frac{1}{2} \frac{\ga_1-b_1}{\ga_0-b_0}\right), \nonumber \\
&& s_2 =  \frac{ 3(-26259 + 123 n_f + 619/3 n^2_f + 144 (9 + 38 n_f)z_3 )}
{384\pi^4 (57 - 2 n_f)(9 - 2 n_f)},\nonumber \\
&&  s_3= \frac{-3} {4976640\pi^6 (57 - 2 n_f) (45 - 2 n_f) (9 - 2 n_f)} \times \nn \\
&& \left(  810 (23404073 + 6156 \pi^4 + 8192244 z_3 - 25485840 z_5)  \right. \nn \\  
&&\left. +216  n_f (-12485305 + 95634 \pi^4 - 31990245 z_3 + 25617600 z_5) \right. \nn \\ 
&&\left. -27  n_f^2 (148085 + 61728 \pi^4 -17165440 z_3 + 6249600 z_5)   \right. \nn \\ 
&&\left. + 6 n_f^3 (817165 + 5472 \pi^4 -1685520 z_3) -146300  n_f^4   \right)
\label{si_exact}
\eea
Alternatively a slightly more involved procedure to determine the $s_i$ is to start from the complete divergent expression (\ref{Fpipapp}), evaluated initially from 
RG invariant bare quantities. One then adds to it an explictly RG invariant perturbative series, constructed in terms
of the bare mass and coupling $m_0$, $g_0$, as 
\be
F^2_{\pi,0}(pert) -\frac{m^2_0}{g_0} \sum_{i\ge 0} h_i \eps^i \equiv F^2_\pi({\rm pert})|_{\rm finite} -\frac{m^2}{g} H(g)\,.
\label{sub0}
\ee
One determines the $h_i$ perturbatively by cancelling the remaining divergences~\cite{gn2}, in such a way that the resulting finite contribution, expressed in terms of
renormalized mass $m_0=Z_m m$ and coupling $g_0=Z_g g\mu^{\eps} $, obeys the homogenous
RG equation (\ref{RGstan}). The $Z_g$, $Z_m$ counterterms in the $\ms$ scheme are easily related to (\ref{beta_conv}), (\ref{ga_conv}) as usual. 
In fact using RG properties the subtraction defined in Eq.~(\ref{sub}) is determined solely by 
the $1/\epsilon$ coefficient in Eq.~(\ref{Fpipapp})~\cite{qcd2}, by solving perturbatively the following equation: 
\be
\left[ n \gamma_m(g)+\beta(g)(\frac{1}{g}-\frac{\partial}{\partial g}) \right] H(g) = -g\,\frac{\partial}{\partial g}\,\left[ \frac{g}{m^2} c_1(g)\right]
\ee
in our normalization, where $n=2$ is the relevant dimension for $F^2_\pi$, $H(g)$ is defined in Eqs.~(\ref{sub}), (\ref{sub0}), and $c_1(g)$ is the coefficient of $1/\epsilon$ in Eqs.~(\ref{Fpipapp}), (\ref{Fpidiv}). 
The two procedures are strictly equivalent, with some simple correspondences between the $h_i$ in (\ref{sub0}) and $s_i$ in (\ref{si_exact}).



\begin{thebibliography}{999}
\bibitem{PrivComm} We kindly thank A.~Maier and P.~Marquard who provided us with the 4-loop coefficient of $\ln m$ and divergent $1/\epsilon$ terms   
for the first $p^2\to 0$ moment of the axial vector correlator, not explicited in \cite{Fpi_4loop}, and useful for consistency crosschecks of RG properties.
\end{thebibliography}
\end{document}